\documentclass[%
 reprint,
superscriptaddress,
 amsmath,amssymb,
 aps,
prb,
]{revtex4-2}

\usepackage{graphicx}
\usepackage{dcolumn}
\usepackage{bm}
\usepackage{braket}
\usepackage{float}
\usepackage{amsmath}
\usepackage{amsfonts}   
\usepackage{amssymb}    
\usepackage{comment}
\renewcommand{\arraystretch}{1.25}
\usepackage{array}
\newcolumntype{P}[1]{>{\centering\arraybackslash}p{#1}}

\begin{document}

\preprint{APS/123-QED}

\title{XY-like Incommensurate Magnetic Order in Ce\textsubscript{2}SnS\textsubscript{5}}

\author{Maximilien F. Debbas}
\affiliation{Department of Nuclear Science and Engineering, Massachusetts Institute of Technology, Cambridge, MA 02139, USA}

\author{Takehito Suzuki}
\affiliation{Department of Physics, Toho University, Funabashi, Japan}

\author{Danielle R. Yahne}
\affiliation{Neutron Scattering Division, Oak Ridge National Laboratory, Oak Ridge, TN 37831, USA}%
 
\author{Joseph G. Checkelsky}%
\affiliation{Department of Physics, Massachusetts Institute of Technology, Cambridge, MA 02139, USA}%

\date{\today}

\begin{abstract}
We report the synthesis of single crystals of Ce\textsubscript{2}SnS\textsubscript{5} through a two-stage chemical vapor transport method. The Ce\textsubscript{2}SnS\textsubscript{5} system is a member of the orthorhombic \textit{Pbam} (No. 55) space group and realizes a distorted trigonal tricapped prism (TTP) crystal field around each cerium site. We characterized the sample through orientation-dependent magnetization and heat capacity measurements to probe the magnetic anisotropy in the system characteristic of XY-like anisotropic Heisenberg model behavior. Ce\textsubscript{2}SnS\textsubscript{5} furthermore enters a zero-field ordered phase under $T_N = 2.4 \, \text{K}$; powder neutron diffraction measurements reveal incommensurate magnetic order near $T_N$. The system then locks into a commensurate, two-\textit{q} magnetic structure below approximately $1.2 \, \text{K}$. This commensurate structure belongs to the Shubnikov group \textit{Pb’a’m’} (MSG 55.359) and realizes the propagation vectors $\vec{q} = (1/3,0,0)$ and $\vec{q} = (0,0,0)$.


\end{abstract}

\maketitle


\section{\label{sec:Introduction} Introduction}

Antiferromagnetic systems such as the MPS\textsubscript{3} family \cite{wildes2017magnetic,PhysRevB.46.5425,grasso2002low} and the BaM\textsubscript{2}(XO\textsubscript{4})\textsubscript{2} \cite{regnault1984magnetic} family provide physical examples of three dimensional systems realizing anisotropic Heisenberg model physics. The magnetic interactions in these systems may be modeled by an anisotropic Heisenberg Hamiltonian with the following general form:

\begin{equation}
    \mathbf{H} = -2 \sum_{i < j} \big( J_\perp \left( \mathbf{S}^x_i \mathbf{S}^x_j + \mathbf{S}^y_{i} \mathbf{S}^y_{j} \right) + J_\parallel \mathbf{S}^z_{i} \mathbf{S}^z_{j}  \big)
\end{equation}

\noindent
where the summation is generally taken over the neighboring sites ${i,j}$ in the lattice, describes the Ising model when $J_\perp = 0$, the XY model when $J_\parallel = 0$, and the isotropic Heisenberg model when $J_\perp = J_\parallel$ \cite{alma990004450580106761}. In real bulk systems, however, the exchange couplings $J_\perp$ and $J_\parallel$ are typically relatively isotropic; anisotropic Ising or XY model physics generally arise from spin-orbit coupling and crystal field effects \cite{alma990004450580106761}. The axial distortion induced by a crystal field modifies the isotropic Heisenberg Hamiltonian to the following form:

\begin{equation}
    \mathbf{H} = -2 J \sum_{i < j}  \vec{\mathbf{S}}_i \cdot \vec{\mathbf{S}}_j  - D \sum_i \big(\mathbf{S}^z_{i}\big)^2
\end{equation}

\noindent
where $D \to +\infty$ describes an Ising system, and $D \to -\infty$ describes an XY system \cite{alma990004450580106761}. Experimentally, the anisotropy of the g-tensor differentiates an Ising system ($g_z \gg g_x,g_y$) from an an XY system ($g_x \approx g_y \gg g_z$) \cite{alma990004450580106761}. We report here that the Ce\textsubscript{2}SnS\textsubscript{5} system may realize XY-like magnetism induced by a highly anisotropic crystal field surrounding the cerium site.

The magnetic behavior of Ce\textsubscript{2}SnS\textsubscript{5} is probed via measurements of the magnetization. This quantity (along with the derived magnetic susceptibility) are obtained from the free energy of the system as:

\begin{gather}
    M_T = - \bigg( \frac{\partial F}{\partial H} \bigg)_T\\
    \chi_T = \bigg( \frac{\partial M}{\partial H} \bigg)_T = - \bigg( \frac{\partial^2 F}{\partial H^2} \bigg)_T
\end{gather}

\noindent
where $M_T$ is the isothermal magnetization, $\chi_T$ is the isothermal magnetic susceptibility, $F$ is the free energy, and $H$ is the applied field strength \cite{alma990004450580106761}. For small applied fields, the magnetic susceptibility may be approximated as $\chi \approx M/H$. 

Ce\textsubscript{2}SnS\textsubscript{5} is a member of the centrosymmetric \textit{Pbam} space group and realizes the La\textsubscript{2}SnS\textsubscript{5} crystal structure. There exists a previous structural report for La\textsubscript{2}SnS\textsubscript{5} on $\sim$100 $\mu$m large single crystals obtained in mapping the the Ln\textsubscript{2}S\textsubscript{3}-SnS\textsubscript{2} (Ln = lanthanide) phase diagram \cite{jaulmes1974structure}. Another literature report on the Ln\textsubscript{2}X\textsubscript{3}-SnX\textsubscript{2} (Ln = lanthanide, X = S or Se) phase diagram also contains some additional structural data for Ce\textsubscript{2}SnS\textsubscript{5} \cite{GUITTARD19761073}.

We report herein a vapor transport method for growing large, high-quality single crystals of Ce\textsubscript{2}SnS\textsubscript{5}. Through orientation-dependent magnetization and heat capacity measurements, we find evidence for a planar ordering of magnetic moments primarily in the \textit{ab}-plane of the crystal. Through powder neutron diffraction, we additionally find evidence for incommensurate order in the zero-field, low-temperature phase. Due to the strong magnetic anisotropy discovered in this system, Ce\textsubscript{2}SnS\textsubscript{5} provides a new opportunity to study XY-like Heisenberg physics in a physical system.

\section{\label{sec:Growth Method} Growth Method}

Single crystals of Ce\textsubscript{2}SnS\textsubscript{5} were synthesized using a two-part growth methodology. The first phase involved reacting the constituent elements at high temperature followed by subsequent regrinding steps to obtain extremely pure Ce\textsubscript{2}SnS\textsubscript{5} powder. The second phase utilized chemical vapor transport (CVT) of this powder with a chloride transport agent. The materials used were 99.9\% pure (metals basis excluding Ta) cerium rod (Thermo Scientific), 99.995\% pure, -100 mesh tin powder (Thermo Scientific), 99.9995\% pure sulfur pieces (Thermo Scientific), and $\geq$99.99\% pure tin (II) chloride (Sigma Aldrich).

\begin{figure}[H]
	\centering 
	\includegraphics[width=1\linewidth]{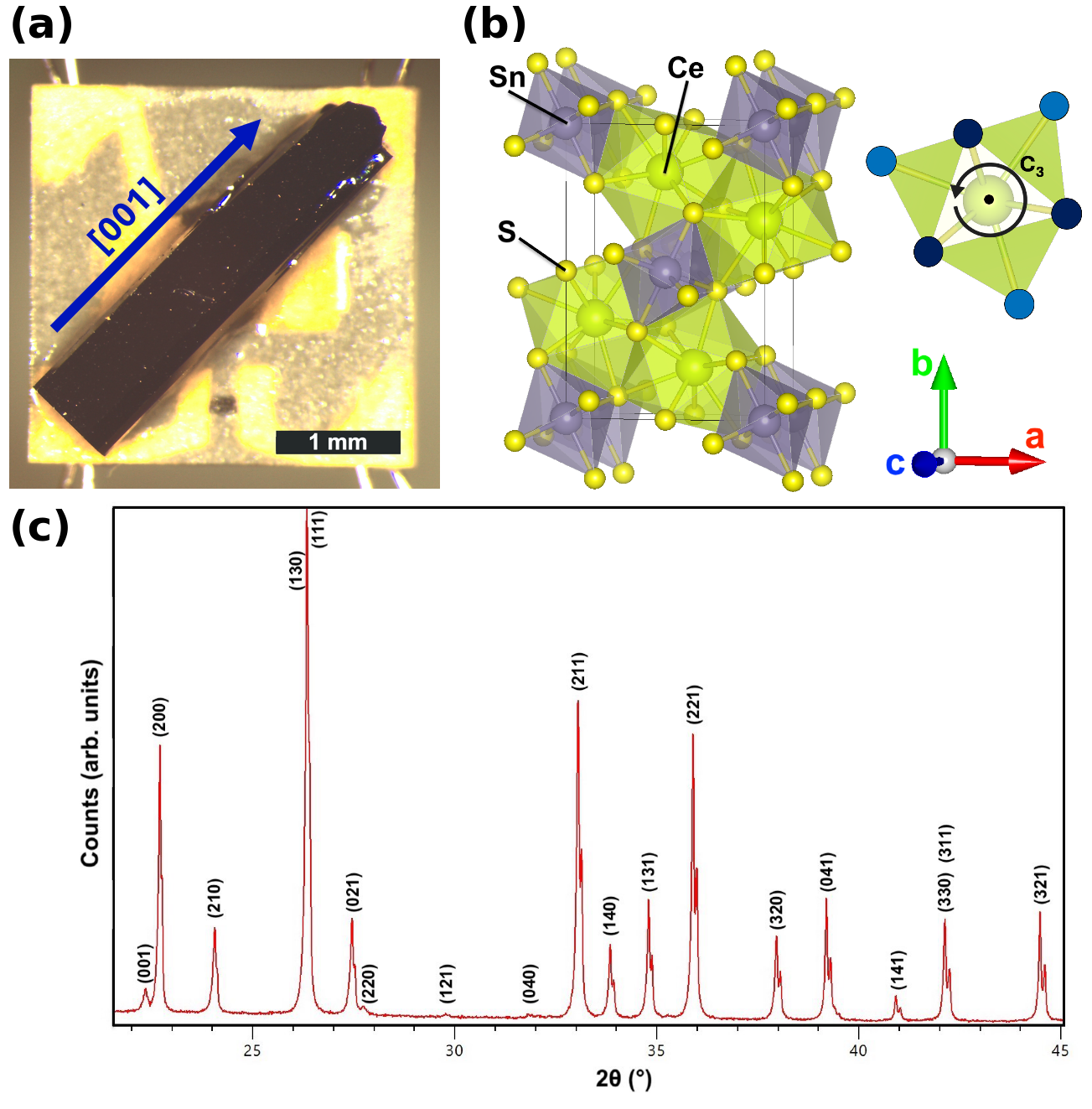}
	\caption{\textbf{(a)} Ce\textsubscript{2}SnS\textsubscript{5} single crystal prepared for a heat capacity measurement with the $[001]$ direction indicated by the blue arrow. \textbf{(b)} Ce\textsubscript{2}SnS\textsubscript{5} crystal structure showing the cerium in green, the sulfur in yellow, and the tin in purple. The coordination polyhedron around each cerium site is shown on the right with the $C_3$ symmetry axis of the undistorted polyhedron labeled and the sulfur atoms forming the polyhedron drawn in blue (the dark blue atoms lie in the plane of the page whereas the light blue atoms lie in a plane behind that of the page). Crystal structure drawn using VESTA \cite{VESTA}. \textbf{(c)} Powder XRD pattern (obtained using copper $K-\alpha$ radiation) of material obtained at the end of the second synthesis phase. The $(hkl)$ indices of the Ce\textsubscript{2}SnS\textsubscript{5} peaks are labeled.}
	\label{fig:CrysStruc}
\end{figure}

For the first phase, a chunk of cerium rod was clipped off and placed into an alumina crucible along with a stoichiometric amount of tin powder and sulfur. This was done under argon in a glovebox, and the alumina crucible containing the starting materials was sealed under high vacuum in a quartz tube. The tube was then heated to 900\textsuperscript{o}C over 36 hours, held at that temperature for 48 hours, cooled to 400\textsuperscript{o}C over a week, and then finally cooled to room temperature over 12 hours. After this initial step, two regrinding steps were performed during which the material from the previous step was ground using a mortar and pestle and resealed in a quartz tube under vacuum. The resealed tube was then heated to 850\textsuperscript{o}C over eight hours, held at that temperature for five days, and then finally cooled to room temperature over eight hours. The purity of the material was checked after each step using powder x-ray diffraction (XRD).

For the second phase, a chloride CVT growth was done using Ce\textsubscript{2}SnS\textsubscript{5} powder from the previous phase and a small amount of tin (II) chloride (0.258 mol\textsubscript{SnCl\textsubscript{2}}/mol\textsubscript{Ce\textsubscript{2}SnS\textsubscript{5}}). The materials were sealed inside a 300 mm long, 12 mm diameter quartz tube under high vacuum and placed in a two-zone tube furnace such that end of the tube containing the starting materials was put at the hot zone and the opposite end was put at the cold zone. The tube was heated up over 12 hours, the hot/cold zones held at 850\textsuperscript{o}C/800\textsuperscript{o}C respectively for 300 hours, and then cooled to room temperature over 12 hours. This was repeated two additional times. At the end of the final stage, several few mm long, prism-shaped crystals of Ce\textsubscript{2}SnS\textsubscript{5} had formed at the hot end. Figure \ref{fig:CrysStruc}(a) shows one of the single crystals obtained from this synthesis batch and Fig. \ref{fig:CrysStruc}(c) shows the powder XRD spectrum of material obtained at the end of the second synthesis phase.

\section{\label{sec:Crystal Structure} Crystal Structure}

The Ce\textsubscript{2}SnS\textsubscript{5} system is a member of the orthorhombic \textit{Pbam} (No. 55) space group. The structure (shown in Figure \ref{fig:CrysStruc}(b) possesses a single mirror symmetry in the \textit{ab}-plane and two half-cell glide symmetries along the \textit{a} and \textit{b} axes. The crystals all realized a prismatic morphology with the $[001]$ direction (\textit{c}-axis) parallel to the long dimension of the prism. This orientation was determined via Laue x-ray diffraction on several single crystals from the synthesis batch. A Rietveld refinement fit of a powder neutron diffraction spectrum taken at $4\, \text{K}$ yielded a a goodness of fit of $\chi^2 = 4.41$ and was used to extract the lattice parameters $a=7.8563(4) \, \text{\AA}$, $b=11.2219(5) \, \text{\AA}$, and $c=3.9563(2) \, \text{\AA}$ which are in reasonable agreement with values reported in the literature \cite{GUITTARD19761073}. Table \ref{T: AtomicPositions} provides the atomic positions of the structure obtained from this fit.


\begin{table}[H]
\caption{\label{T: AtomicPositions}Atomic positions and isotropic displacement parameters ($B_{iso}$) for Ce\textsubscript{2}SnS\textsubscript{5} obtained from the Rietveld refinement of a $4\, \text{K}$ powder neutron diffraction spectrum.}
\begin{ruledtabular}
\begin{tabular}{p{0.5cm} P{1cm} P{1cm} P{1cm} P{1cm} P{1cm}}
Site & Wyckoff & $x/a$ & $y/b$ & $z/c$ & $B_{iso}$ \\ [0.8ex]
\hline
Ce1 & $4h$ & 0.0699(2) & 0.3309(2) & $\tfrac{1}{2}$ & 0.68(8) \\ [0.6ex]
Sn1 & $2a$ & 0 & 0 & 0 & 0.37(8) \\ [0.6ex]
S1 & $4h$  & 0.1860(4) & 0.0729(3) & $\tfrac{1}{2}$ & 0.39(11) \\ [0.6ex]
S2 & $4g$  & 0.3536(4) & 0.3004(3) & 0 & 0.31(11) \\ [0.6ex]
S3 & $2c$  & 0 & $\tfrac{1}{2}$ & 0 & 0.81(11) \\ [0.6ex]
\end{tabular}
\end{ruledtabular}
\end{table}

Each cerium atom coordinates with nine nearest neighbor sulfur atoms which form a slightly distorted tricapped trigonal prism (TTP) coordination polyhedron (shown to the right of Fig. \ref{fig:CrysStruc}(b)). Were it undistorted, this coordination polyhedron would realize the point group $D_{3h}$ with the $C_3$ rotational axis of the polyhedron aligned parallel to the \textit{c}-axis. The distortion may be considered in two ``steps'' as $D_{3h} \to C_{2v} \to C_s$. The $D_{3h} \to C_{2v}$ step accounts for the vast majority of the distortion effects as the $C_{2v} \to C_s$ step is slight. The coordination polyhedron is well-modeled by an ionic crystal field where the sulfur realizes S\textsuperscript{2-} and the cerium realizes Ce\textsuperscript{3+}.

\subsection{\label{Crystal Field Splitting} Crystal Field Splitting}

The coordination polyhedron formed by the nine S\textsuperscript{2-} ions causes crystal field splitting of the localized $4f$-electron of the Ce\textsuperscript{3+} ion at the center of the polyhedron. Due to strong spin-orbit coupling in rare earth systems, the single $4f$-electron couples to the orbital angular momentum to yield a sixfold degenerate ground state multiplet $^2 F_{5/2}$ and an eightfold degenerate excited state multiplet $^2 F_{7/2}$. The Ce\textsuperscript{3+} free-space spin-orbit coupling between these two multiplets of $3150 \, \text{K}$ is far larger than both the crystal field splitting and laboratory temperature scales \cite{10.1093/oso/9780198520276.003.0001}. The excited state multiplet $^2 F_{7/2}$ may thereby be ignored.

As the ground state multiplet $^2 F_{5/2}$ is characterized by the half-quantized total angular momentum $J=5/2$, the crystal field splitting must be analyzed using the crystal double group of the coordination polyhedron due to the odd parity of $^2 F_{5/2}$ under a $2\pi$ rotation \cite{Tinkham}. The $^2 F_{5/2}$ multiplet is then subjected to the double group $D_{3h}'$ in the undistorted crystal field such that the $^2 F_{5/2}$ multiplet breaks up into the three, doublet irreducible representations $\Gamma_7 \oplus \Gamma_8 \oplus \Gamma_9$. 

The first distortion step lowers the symmetry to the double group $C_{2v}'$ such that the $^2 F_{5/2}$ multiplet breaks up into three copies of the $\Gamma_{5}$ Kramers' doublet (\textit{i.e.}, $3\Gamma_{5}$). Should the distortion be considered in its entirety, the symmetry would formally be lowered to the double group $C_{s}'$ such that the $^2 F_{5/2}$ multiplet breaks up into three copies of the $\Gamma_{3,4}$ Kramers' doublet (\textit{i.e.} $ 3\Gamma_{3,4}$). Both distortions preserve the energy level structure as they simply shift the energies of the Kramers' doublets. The level structure of Ce\textsuperscript{3+} may then be modeled as a three level system of doublets even when distortion of the crystal field is considered.

\subsection{\label{Magnetic Anisotropy} Magnetic Anisotropy}

As the distortion is slight, the Kramers' pairs will be considered here for the undistorted $D_{3h}'$ double group. The generic crystal electric field (CEF) potential for a TTP ligand field is given by Eq. (\ref{Eq: TTP}):

\begin{gather}
    \label{Eq: TTP} V_{TTP} = B_0^2 C_0^2 + B_0^4 C_0^4 + B_0^6 C_0^6 + B_6^6 \left( C_{-6}^6 + C_6^6 \right)\\
    \label{Eq: Racah} C_q^k (\theta,\phi) = \sqrt{\frac{4\pi}{2k+1}} Y_k^q (\theta,\phi)
\end{gather}

\noindent
where $B_q^k$ are the crystal field parameters, and $C_q^k (\theta,\phi)$ are tensor operators (related to the spherical harmonics $Y_k^q (\theta,\phi)$ via Eq. (\ref{Eq: Racah})) \cite{RareEarth23}. The polar angles $\theta,\phi$ are defined relative to the $C_3$ rotation axis of the crystal field, \textit{i.e.}, along the $c$-axis of the crystal. Treating $V_{TTP}$ as a perturbation and diagonalizing it in the $\ket{J=5/2,m_J}$ basis yields the following degenerate pairs of wavefunctions:

\begin{gather}
    \label{Eq: 1/2} \Gamma_7 \quad : \quad \ket{\psi_{\pm1/2}} = \ket{5/2,\pm 1/2}\\
    \label{Eq: 3/2} \Gamma_8 \quad : \quad  \ket{\psi_{\pm3/2}} = \ket{5/2,\pm 3/2}\\
    \label{Eq: 5/2} \Gamma_9 \quad : \quad  \ket{\psi_{\pm5/2}} = \ket{5/2,\pm 5/2}
\end{gather}

\noindent
each corresponding to a Kramers' doublet in the TTP energy level structure. Computing the g-tensor associated with each of these degenerate subspaces (see Appendix \ref{sec: Magnetization of the Kramers' Doublets}) yields:

\begin{gather}
    \label{Eq: g1/2}\ket{\psi_{\pm1/2}} \qquad : \qquad g_\parallel = \frac{6}{7} \quad g_\perp = \frac{18}{7} \\
    \label{Eq: g3/2}\ket{\psi_{\pm3/2}} \qquad : \qquad g_\parallel = \frac{18}{7} \quad g_\perp = 0 \\
    \label{Eq: g5/2}\ket{\psi_{\pm5/2}} \qquad : \qquad g_\parallel = \frac{30}{7} \quad g_\perp = 0
\end{gather}

\noindent
which highlights the magnetic anisotropy of the system compared to the isotropic Landé g-factor of $g_{J=5/2} = 6/7$. Including the effects of the distortion will mix other components of $\ket{J=5/2,m_J}$ into the wavefunctions spanning each degenerate subspace and will result in nonzero values of $g_\parallel$ and $g_\perp$ for each of the Kramers' doublets.

\section{\label{sec:Measurements} Measurements}

Herein, we present results derived from magnetization and heat capacity measurements of three Ce\textsubscript{2}SnS\textsubscript{5} single crystals (\textit{A}, \textit{B}, and \textit{C}) which were grown in the same synthesis batch. Note that the Ce\textsubscript{2}SnS\textsubscript{5} system was found to be an electrical insulator. We also present results from a zero-field powder neutron diffraction measurement performed at Oak Ridge National Laboratory.

\subsection{Magnetization}
\label{Magnetization}

Magnetization $M$ was measured using SQUID magnetometry in a Quantum Design MPMS3 system. Single crystals were affixed to a quartz rod using GE varnish and DC magnetization measurements were performed using Vibrating Sample Magnetometer (VSM) mode.

\begin{figure}[H]
	\centering 
	\includegraphics[width=1\linewidth]{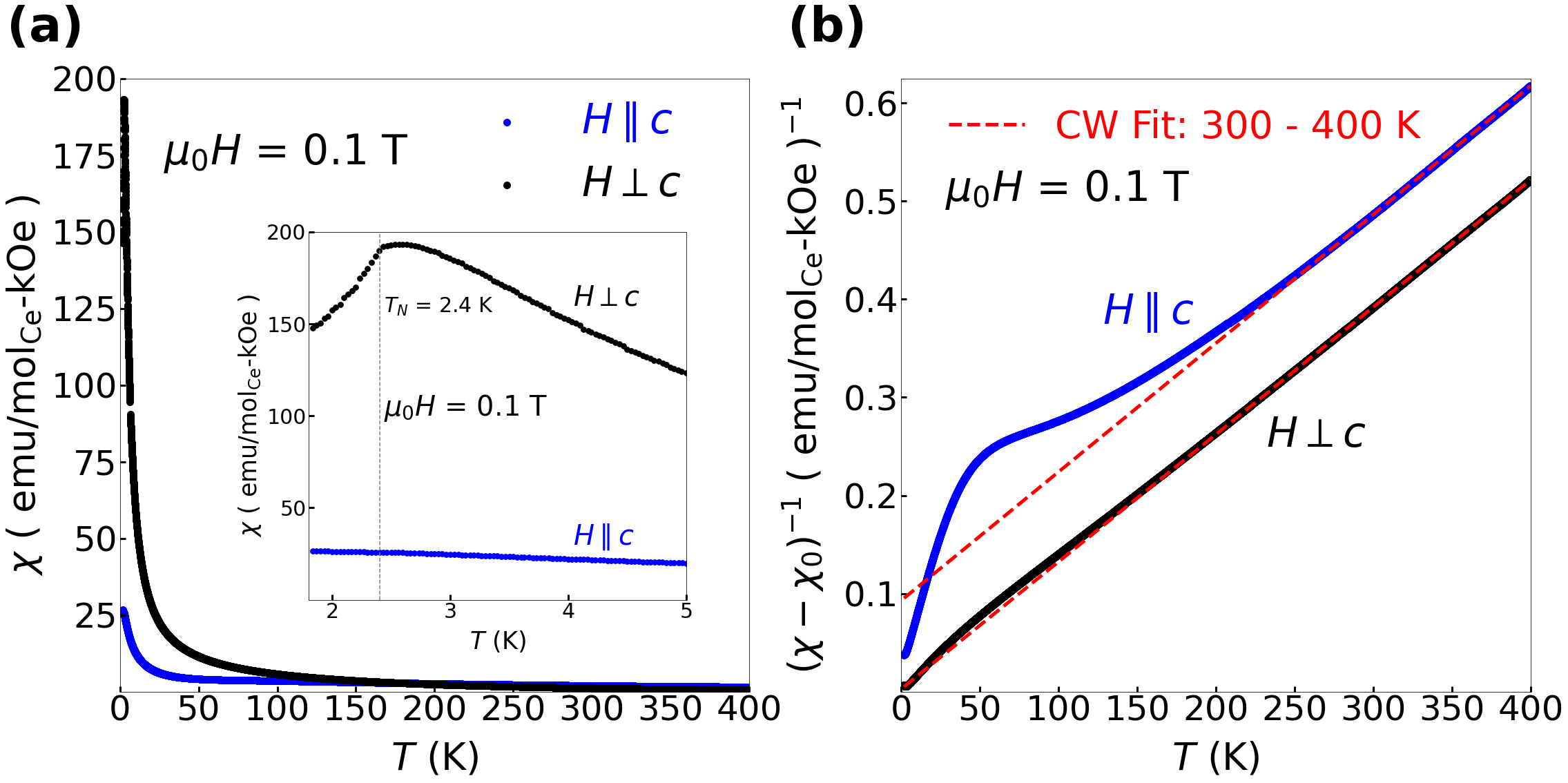}
	\caption{\textbf{(a)} Magnetic susceptibility of crystal \textit{A} taken in $0.1 \, \text{T}$ of applied magnetic field with the field aligned perpendicular to the \textit{c}-axis (black) and parallel to the \textit{c}-axis (blue). The inset shows the low temperature magnetic susceptibility with the magnetic transition at $T_N = 2.4 \, \text{K}$ marked by the vertical dashed line. \textbf{(b)} Inverse magnetic susceptibility from the same measurement with a Curie-Weiss fit between $300-400 \, \text{K}$ (dashed, red curve).}
	\label{fig:mag}
\end{figure}

Figure \ref{fig:mag}(a) shows the magnetic susceptibility of Ce\textsubscript{2}SnS\textsubscript{5} with the field aligned perpendicular to the \textit{c}-axis (black) and parallel to the \textit{c}-axis (blue). The inset shows the low temperature magnetic susceptibility; a peak is visible at $T_N = 2.4 \, \text{K}$ when the field is aligned perpendicular to the \textit{c}-axis, and a corresponding kink appears when the field is aligned parallel to the \textit{c}-axis. These features indicate the onset of a magnetic phase transition.

Figure \ref{fig:mag}(b) shows the inverse magnetic susceptibility (with the $T$-independent offset $\chi_0$ subtracted from the susceptibility) for both orientations of the crystal. The data was fit between $300-400 \, \text{K}$ (dashed, red curve) by a Curie-Weiss model with a constant offset via the following equation:

\begin{equation}\label{eq:X}
    \chi = \frac{1}{T-\theta_{CW}} \frac{N_A}{3k_B} \mu_{eff}^2 + \chi_0
\end{equation}

\noindent
where $\chi$ is the magnetic susceptibility per mole, $\theta_{CW}$ is the Curie temperature, $N_A$ Avogadro's number, $\chi_0$ is the $T$-independent susceptibility component, and $\mu_{eff}$ is the effective moment in units of $\mu_B$. The parameters obtained from this fit are displayed in Table \ref{T: CW}. The effective moments for both orientations show good agreement with the free-space effective Ce\textsuperscript{3+} moment of $2.54 \, \mu_B$. The high temperature Curie-Weiss fit of the data deviates strongly under $200 \, \text{K}$ due to the strong effects of the crystal field in this orientation.

\begin{table}[H]
\caption{\label{T: CW}Parameters from the Curie-Weiss fit of Ce\textsubscript{2}SnS\textsubscript{5} crystal $A$ between 300 K - 400 K in a field of $\mu_0 H = 0.1 \, \text{T}$ aligned both perpendicular and parallel to the \textit{c}-axis.}
\begin{ruledtabular}
\begin{tabular}{p{3cm} P{2cm} P{2cm}}
  Parameter &  $H \perp c$ &  $H \parallel c$   \\ [0.8ex] 
  \hline
  $\theta_{CW}$ (K) & $-2.1 \pm 0.5$ & $-70.9 \pm 0.5$ \\ [0.6ex] 
  $\mu_{eff}$ ($\mu_B$/Ce) & $2.483 \pm 0.003$ & $2.470 \pm 0.003$ \\ [0.6ex] 
\end{tabular}
\end{ruledtabular}
\end{table}

\begin{figure}[H]
	\centering 
	\includegraphics[width=1\linewidth]{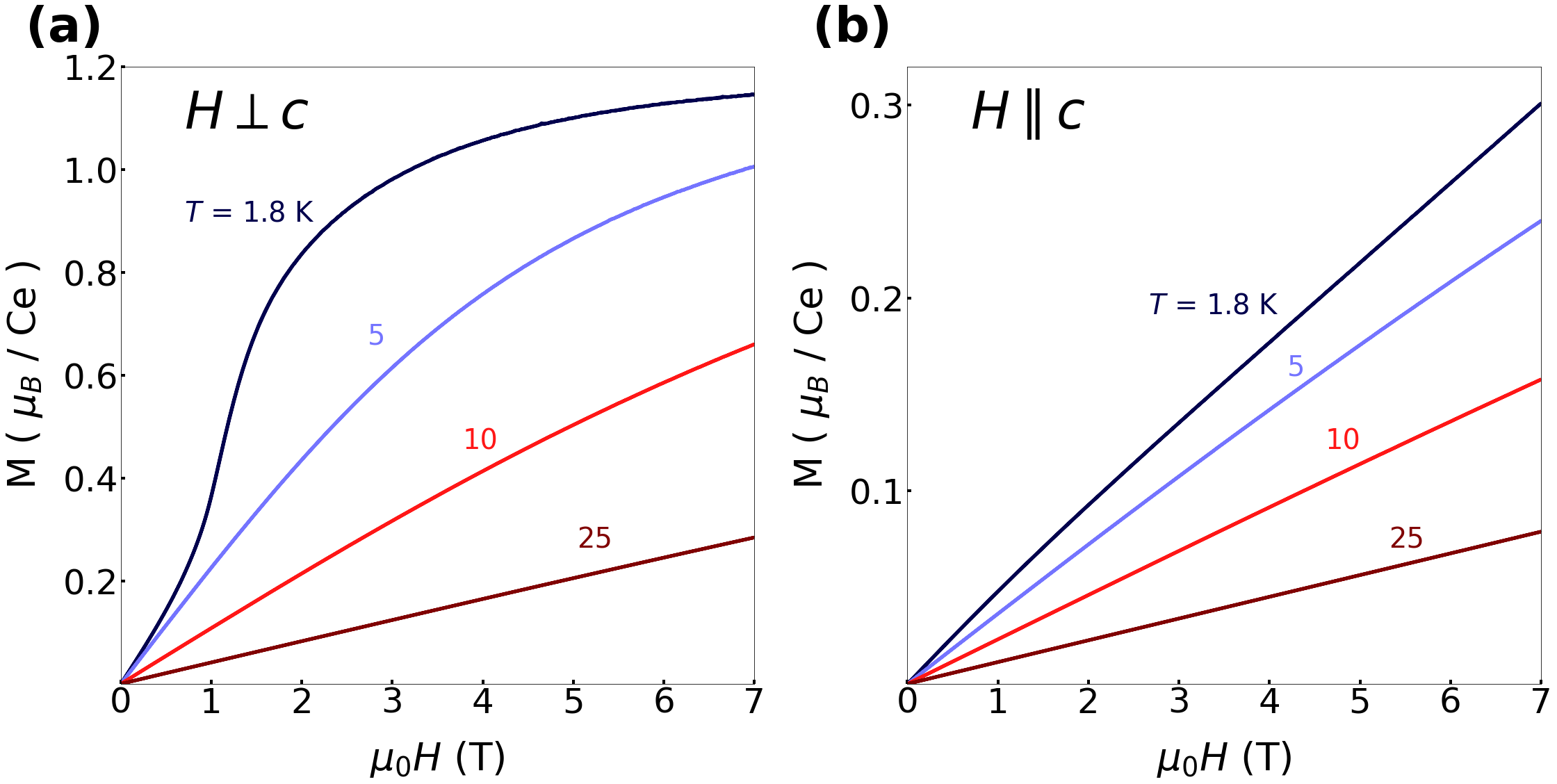}
	\caption{Magnetization of crystal \textit{A} taken as a function of applied magnetic field perpendicular to the \textit{c}-axis \textbf{(a)} and parallel to the \textit{c}-axis \textbf{(b)}. The magnetization was measured at several temperatures both below and above $T_N = 2.4 \, \text{K}$ and is presented in units of $\mu_B$ per cerium atom.}
	\label{fig:MvH}
\end{figure}

Figure \ref{fig:MvH} shows the magnetization as a function of applied magnetic field both perpendicular and parallel to the \textit{c}-axis. When the field is applied perpendicular to the \textit{c}-axis (at $T = 1.8 \, \text{K}$, below $T_N$) the magnetization reaches $1.15 \, \mu_B$ per cerium atom at $\mu_0 H = 7 \, \text{T}$ and appears to saturate. At $T = 1.8 \, \text{K}$, the magnetization also exhibits an inflection point near $\mu_0 H = 1 \, \text{T}$; this inflection point disappears for temperatures above $T_N$ whereat the magnetization follows a standard Brillouin function. When the field is applied parallel to the \textit{c}-axis, the magnetic susceptibility remains linear up to $\mu_0 H = 7 \, \text{T}$ for all temperatures.

The low temperature magnetization was fit to the expected magnetization of a single Kramers' doublet (see Appendix \ref{sec: Magnetization of the Kramers' Doublets}) to yield ground state g-factors of $g_\parallel = 1.2$ and $g_\perp = 2.5$ which are very close to the values predicted for the
ground state spanned by $\ket{\psi_{\pm1/2}}$. The experimentally determined ground state value of $g_\parallel$ is larger than the expected value of $6/7$ for this doublet due to thermal population of the excited Kramers' doublets mixing in larger, nonzero values of $g_\parallel$. The expected saturation magnetizations of the $\ket{\psi_{\pm1/2}}$ doublet are $M_s^\parallel = \frac12 g_\parallel = 3/7$ and $M_s^\perp = \frac12 g_\perp = 9/7$; these values are consistent with the behavior of the magnetization seen in Fig. \ref{fig:MvH}.

\subsection{Heat Capacity}
\label{Heat Capacity}

Heat capacity was measured using a Quantum Design PPMS system with the heat capacity module enabled. Single crystals were affixed to heater platform using Apiezon N grease. A background heat capacity measurement was taken of the grease prior to mounting the crystal such that it could be subtracted from the total heat capacity after the crystal was mounted.

\begin{figure}[H]
	\centering 
	\includegraphics[width=1\linewidth]{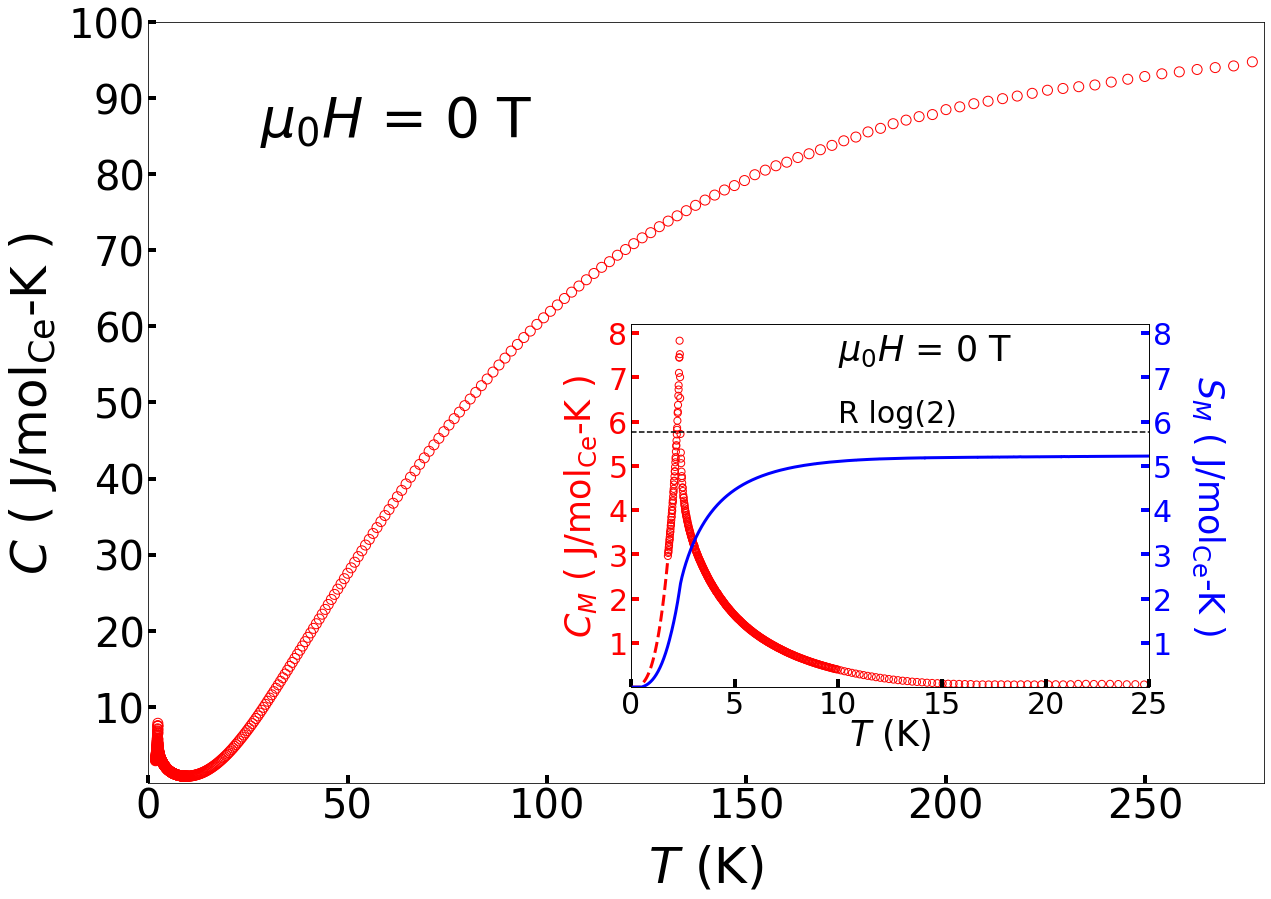}
	\caption{Heat capacity of crystal \textit{B} taken in zero field showing the magnetic transition at $T_N = 2.4 \, \text{K}$. The inset shows the magnetic part of the heat capacity in red and the computed magnetic entropy in blue below $25 \, \text{K}$. The magnetic part of the heat capacity extrapolated to zero is shown as a dashed red line. The magnetic entropy lost through the transition is consistent with the doublet ground state of the cerium $4f$-electron level structure.}
	\label{fig:HCvT}
\end{figure}

Figure \ref{fig:HCvT} shows the heat capacity of Ce\textsubscript{2}SnS\textsubscript{5} in zero field. This data was fit above $25 \, \text{K}$ with a combined Debye and Einstein phonon model (see Appendix \ref{sec: Fitting the Heat Capacity} for more details) such that the extrapolated fit could serve as a phonon background under $25 \, \text{K}$. The inset of Fig. \ref{fig:HCvT} shows the magnetic heat capacity $C_M$ obtained by subtracting this phonon background from the data in this temperature regime. The data below the $T_N = 2.4 \, \text{K}$ transition was fit by a $C_M \sim T^\alpha$ power law to obtain the exponent $\alpha = 3.03$ which is consistent with the heat capacity of antiferromagnetic magnons in three dimensions.

This power law fit was used to extrapolate the data down to zero temperature and provide a continuous magnetic heat capacity function for computing the magnetic entropy $S_M$. The magnetic entropy saturates to a value of $R \log{\Omega}$ with $\Omega = 1.87$ which agrees with the predicted doublet ground state of the Ce\textsuperscript{3+} level structure. 

\begin{figure}[H]
	\centering 
	\includegraphics[width=1\linewidth]{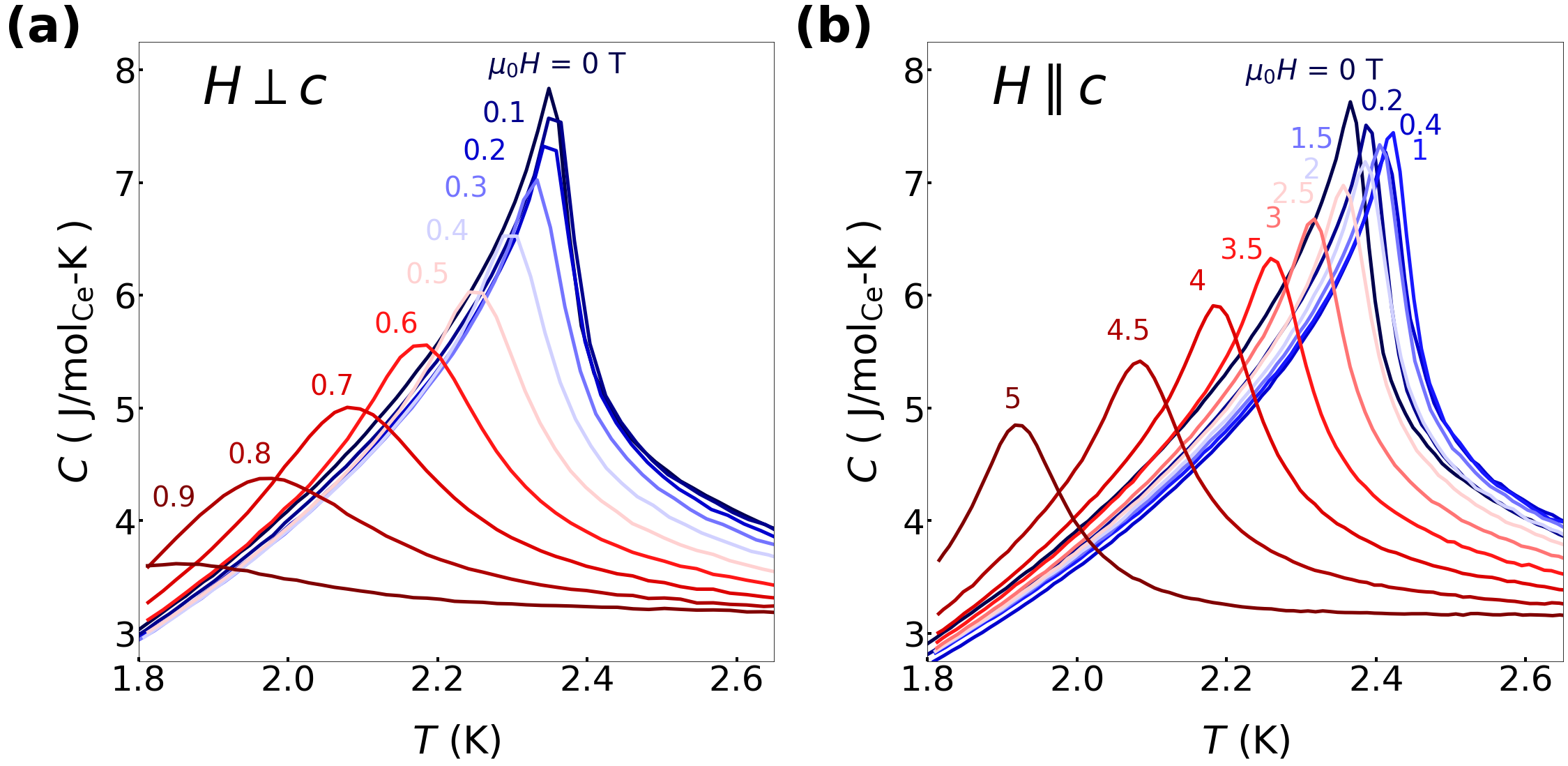}
	\caption{Heat capacity measured around the magnetic transition for various applied magnetic fields. \textbf{(a)} Heat capacity of crystal \textit{B} with field aligned perpendicular to the \textit{c}-axis. \textbf{(b)} Heat capacity of crystal \textit{C} with field aligned parallel to the \textit{c}-axis.}
	\label{fig:HCvT_field}
\end{figure}

Figure \ref{fig:HCvT_field} shows the evolution of the magnetic transition in increasing magnetic fields applied both perpendicular and parallel to the \textit{c}-axis. For field applied perpendicular to the \textit{c}-axis, the peak persists above $1.8 \, \text{K}$ until at least $\mu_0 H = 0.9 \, \text{T}$. For field applied parallel to the \textit{c}-axis persists above $1.8 \, \text{K}$ until at least $\mu_0 H = 5 \, \text{T}$.

For both orientations of the crystal relative to the applied magnetic field, heat capacity measurements were taken up to $9 \, \text{T}$ to reveal a large Schottky anomaly. This Schottky anomaly arises from the Ce\textsuperscript{3+} electronic level structure, as $^{140}\text{Ce}$ (the primary isotope in natural cerium) does not possess a nuclear moment and thereby does not exhibit a nuclear Schottky anomaly. The behavior of the Schottky anomaly for both orientations of the crystal is discussed further in Appendix \ref{sec: The Schottky Anomaly}.

\subsection{Phase Diagram}
\label{Phase Diagram}

\begin{figure}[H]
	\centering 
	\includegraphics[width=1\linewidth]{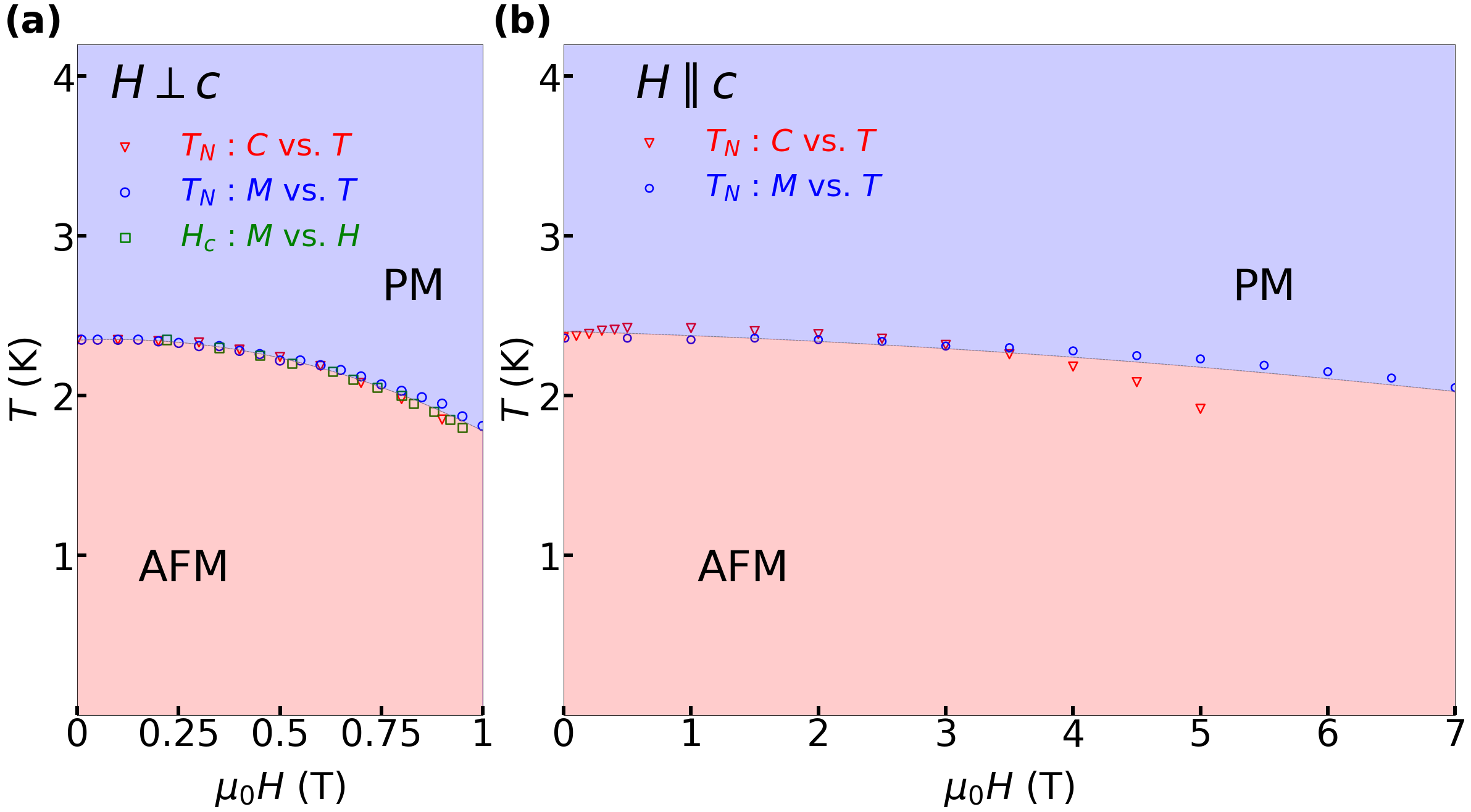}
	\caption{Phase diagram for Ce\textsubscript{2}SnS\textsubscript{5} with field applied perpendicular \textbf{(a)} and parallel \textbf{(b)} to the \textit{c}-axis showing the boundary between antiferromagnetism (AFM) and paramagnetism (PM). The data is derived from the position $T_N$ of the features in heat capacity (red triangle) and magnetization (blue circle) measured in fixed field, and the position of the kink at $H_C$ in magnetization (green square) measured at fixed temperature.}
	\label{fig:PD}
\end{figure}

The anisotropy in the magnetic ordering of Ce\textsubscript{2}SnS\textsubscript{5} is illustrated in the phase diagram shown in Figure \ref{fig:PD} derived from heat capacity and magnetization data with field applied both perpendicular and parallel to the \textit{c}-axis. The phase boundaries could only be experimentally measured down to $1.8 \, \text{K}$, but an upper bound on the critical field corresponding to $T=0$ can be determined by the field at which the Schottky anomaly emerges above $1.8 \, \text{K}$.

Figure \ref{fig:PD}(a) shows the phase boundary when field is applied perpendicular to the \textit{c}-axis. It derives the transition temperature $T_N$ from the position of the peak in heat capacity (red triangle marker) and magnetization (blue circle marker) measured in fixed field. It also plots the critical field $H_C$ (green square marker) corresponding to the kink in magnetization measured at fixed temperature. The value of $H_C$ was determined from the position of the extremum of the second derivative of magnetization with respect to field. 

Figure \ref{fig:PD}(b) describes the phase boundary when field is applied parallel to the \textit{c}-axis. It contains values of $T_N$ derived from the the position of the peak heat capacity  (red triangle marker) as well as the kink in the magnetization (blue circle marker) measured in fixed field. The location of the kink in magnetization was determined from the position of the extremum of the first derivative of magnetization with respect to temperature.

\subsection{Powder Neutron Scattering}
\label{Powder Neutron Scattering}

Powder neutron diffraction was performed at Oak Ridge National Laboratory using the HB-2A diffractometer at the High Flux Isotope Reactor (HIFR). Four grams of powdered Ce\textsubscript{2}SnS\textsubscript{5} were sealed under helium in an aluminum sample can and a $^3\text{He}$ cryostat was used to cool the sample under $2 \, \text{K}$. Neutron scattering spectra ($\lambda = 2.41 \, \text{\AA}$) were taken in zero field at several temperatures between $300 \, \text{mK}$ and  $4 \, \text{K}$.

\begin{figure}[H]
	\centering 
	\includegraphics[width=1\linewidth]{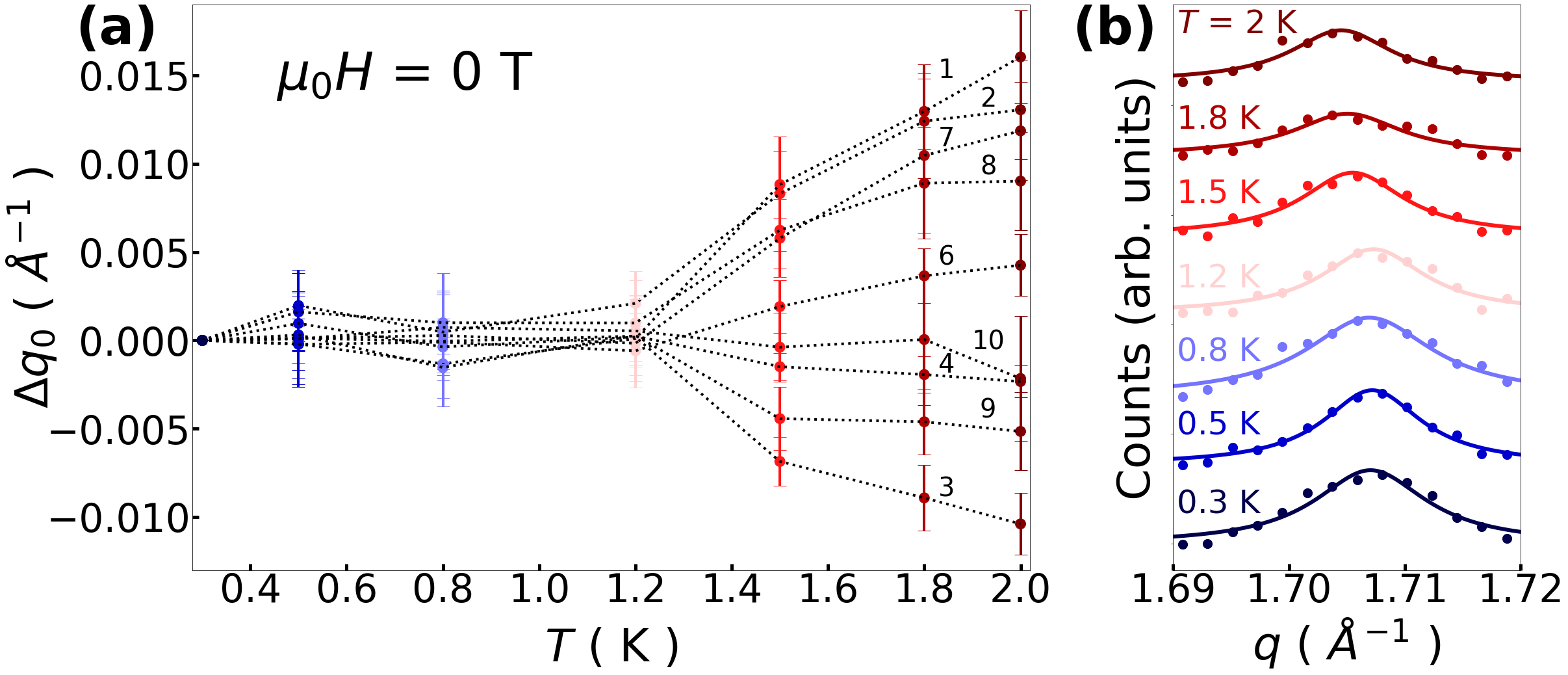}
	\caption{\textbf{(a)} The shift $\Delta q_0 = q_0 - q_0^*$ of the position $q_0$ of selected magnetic reflections at a fixed temperature relative to their positions $q_0^*$ at $T = 300 \, \text{mK}$. The individual traces are numbered according the the notational scheme outlined in Appendix \ref{sec: Fitting Selected Reflection Peaks}. \textbf{(b)} Temperature evolution of the fourth magnetic reflection peak showing the shifting behavior under $T_N$ before locking into a fixed position under $1.2 \, \text{K}$. Note that these data have been offset vertically.}
	\label{fig:MagRefEv}
\end{figure}

Magnetic reflection peaks appear in the neutron diffraction spectra taken below the $T_N = 2.4 \, \text{K}$; a $4 \, \text{K}$ spectrum provided a high temperature background to subtract from spectra showing magnetic reflection peaks. Below $2.4 \, \text{K}$, the magnetic reflection peaks appear to shift until they lock into fixed positions at approximately $1.2 \, \text{K}$. This shifting behavior is shown in Fig. \ref{fig:MagRefEv}(a) which contains a plot of the shift $\Delta q_0$ of selected magnetic reflection peaks relative to their positions at $T = 300 \, \text{mK}$. These magnetic reflection peaks were found to correspond to a propagation vector of $\vec{q} = (1/3,0,0)$. Figure \ref{fig:MagRefEv}(b) explicitly shows the shifting behavior for the fourth magnetic reflection peak ($(030)+(1/3, 0, 0)$ at $300 \, \text{mK}$). The shifting of the magnetic reflection peaks as the system is cooled down to $1.2 \, \text{K}$ suggests the presence of incommensurate order just below $T_N$ which eventually locks into commensurate order. Appendix \ref{sec: Fitting Selected Reflection Peaks} presents additional details on the fitting of these selected magnetic reflections.

\begin{figure}[H]
	\centering 
	\includegraphics[width=1\linewidth]{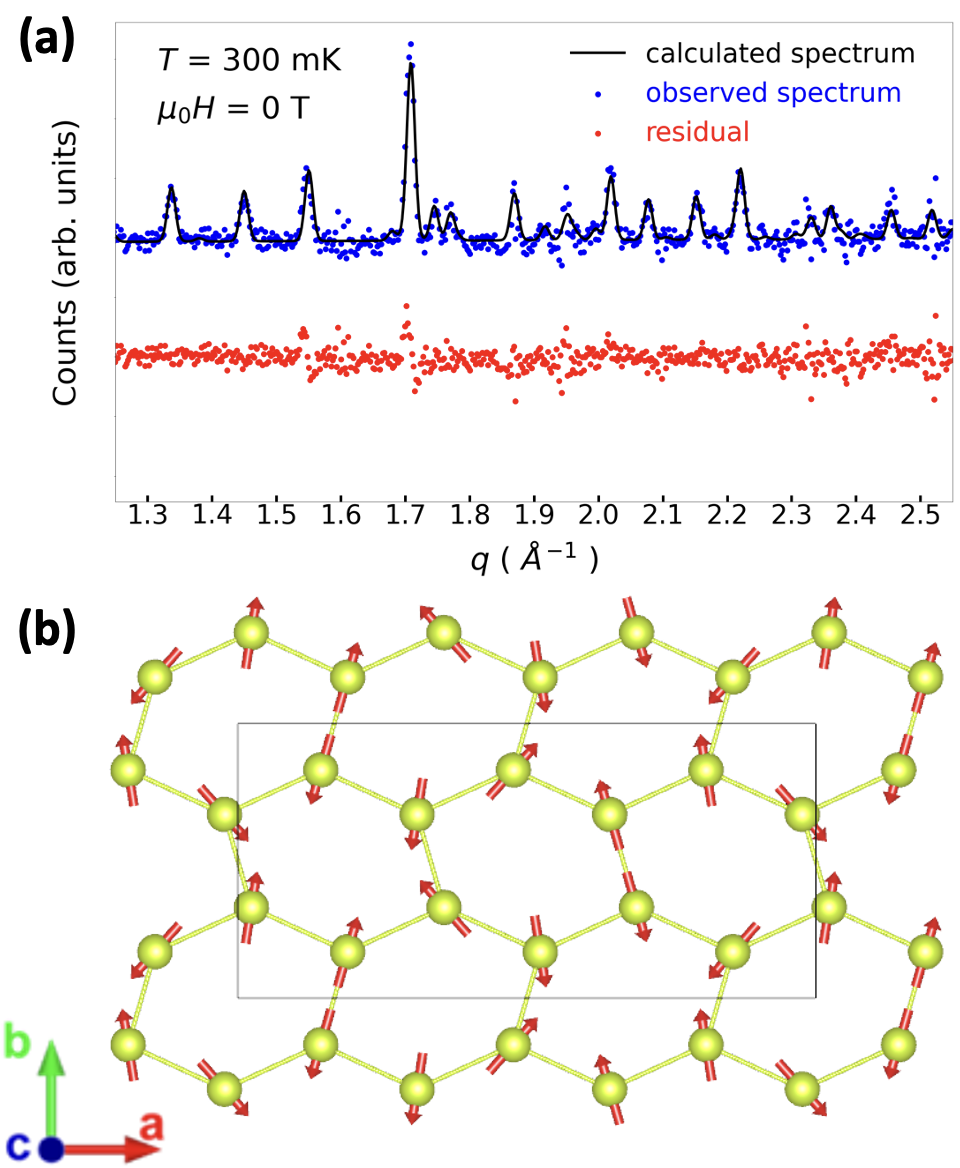}
	\caption{\textbf{(a)} Zero-field, background-subtracted powder neutron diffraction spectrum ($\lambda = 2.41 \, \text{\AA}$) of Ce\textsubscript{2}SnS\textsubscript{5} at $300 \, \text{mK}$. The observed spectrum is shown in blue, the calculated spectrum is shown as a black line, and the residual is shown in red. The raw $4 \, \text{K}$ and $300 \, \text{mK}$ neutron diffraction data are shown in Fig. \ref{fig:403} in Appendix \ref{sec: Fitting Selected Reflection Peaks}. \textbf{(b)} The low-temperature, commensurate magnetic structure of Ce\textsubscript{2}SnS\textsubscript{5}. The red arrows indicate the direction of the magnetic moment on each cerium site and the sulfur and tin atoms have been removed to better illustrate the 120\textsuperscript{o}-like magnetic structure on a distorted honeycomb lattice of cerium atoms in the \textit{ab}-plane. Crystal structure drawn using VESTA \cite{VESTA}.}
	\label{fig:359_struct}
\end{figure}

A Rietveld refinement of the background-subtracted $300 \, \text{mK}$ magnetic reflection spectrum was performed using the FullProf software suite to fit the magnetic structure of the low-temperature commensurate phase \cite{FullProf}. Subject to the constraint that the magnetic moments on the cerium sites have the same magnitude, the best fitting magnetic structure belongs to the Shubnikov group \textit{Pb’a’m’} (MSG 55.359) and adopts a structure wherein all moments point in the \textit{ab}-plane. The observed diffraction spectrum, calculated spectrum, and residual are shown in Fig. \ref{fig:359_struct}(a) and the refined magnetic structure is shown in Fig. \ref{fig:359_struct}(b). The magnetic structure is reminiscent of a 120\textsuperscript{o}-like structure on a distorted honeycomb lattice of cerium atoms similar to that of EuCl\textsubscript{2} within a layer \cite{Kramer2000Triangular}. MSG 55.359 is furthermore compatible with an additional $\vec{q} = (0,0,0)$ irreducible representation which suggests a 2-\textit{q} magnetic structure. Additional details regarding this analysis are presented in Appendix \ref{sec: Fitting The Magnetic Structure}.

\section{Discussion}
\label{sec:Discussion}

The Ce\textsubscript{2}SnS\textsubscript{5} system realizes highly anisotropic magnetic order at low temperatures in a reduced symmetry environment provided by the distorted TTP crystal field surrounding each cerium site. From measurements of the perpendicular and parallel components of the g-tensor, the ground state is described by the $\Gamma_7$ Kramers' doublet in the TTP crystal electric field. In this environment, the magnetic order appears to be more resilient to magnetic fields applied parallel to the $c$-axis compared to fields applied perpendicular to the $c$-axis. We hypothesize that this behavior arises from XY-like anisotropy which biases the magnetic moments to point in the \textit{ab}-plane.

Such a magnetic ordering is supported by the magnetization response illustrated in Figure \ref{fig:MvH}. Due to the strong in-plane anisotropy, when the applied magnetic field is oriented perpendicular to the $c$-axis, the moments point in the same plane as the field such that it drives the system out of the ordered phase above approximately $1 \, \text{T}$ and allows the magnetization to saturate by $7 \, \text{T}$. When the field is oriented near parallel to the $c$-axis, the moments point almost orthogonally to it such that increasing the field only serves to cant the moments out of the \textit{ab}-plane; this effect is weak enough not to saturate the magnetization even by $7 \, \text{T}$ at $1.8 \, \text{K}$ in this orientation. Above the ordering temperature, the ratio $g_\perp/g_\parallel$ was found to be approximately two (see Appendix \ref{sec: Magnetization of the Kramers' Doublets}). This anisotropy in the moment induced by the crystal field suggests XY-like behavior for temperatures $T>T_N$ in the paramagnetic phase. 

The behavior of the magnetic susceptibility below $T<T_N$ (see inset of Fig. \ref{fig:mag}(a)) suggests that the system realizes antiferromagnetic order in the \textit{ab}-plane. Powder neutron diffraction reveals magnetic reflection peaks that shift as the temperature decreases below $T_N$ until they eventually lock into fixed positions below approximately $1.2 \, \text{K}$. This behavior suggests the presence of incommensurate order just below $T_N$ before the system locks into a commensurate magnetic structure upon cooling. The best fitting low-temperature commensurate magnetic structure belongs to the Shubnikov group \textit{Pb’a’m’} (MSG 55.359) and adopts a two-\textit{q} 120\textsuperscript{o}-like magnetic structure on a distorted honeycomb lattice of cerium atoms in the \textit{ab}-plane with the magnetic moments confined to point in the \textit{ab}-plane. Additional neutron diffraction measurements on single crystals would be an important future research direction to elucidate the nature of the incommensurate magnetic structure.

\section{\label{sec:Conclusion} Conclusion}

The presented data shows that the Ce\textsubscript{2}SnS\textsubscript{5} system exhibits strong in-plane magnetic anisotropy arising from the TTP crystal field surrounding the cerium site. In the paramagnetic phase, the system is characterized by an anisotropic g-tensor closely matching the expected magnetic behavior of the ground state $\Gamma_7$ Kramers' doublet. The system realizes magnetic order upon cooling below $T_N = 2.4 \, \text{K}$ and the magnetic entropy associated with the zero-field peak in heat capacity affirms the doublet nature of the ground state. Powder neutron diffraction shows that the zero-field phase exhibits incommensurate order below $T_N$ that locks into a commensurate magnetic structure below $1.2 \, \text{K}$. The low-temperature commensurate structure was found to be a two-\textit{q} 120\textsuperscript{o}-like magnetic structure on a distorted honeycomb lattice of cerium atoms in the \textit{ab}-plane.

The strong magnetic anisotropy in the system suggests that it realizes three dimensional XY model physics. Ce\textsubscript{2}SnS\textsubscript{5} provides a simple material platform to study the interplay between incommensurate order and the XY model. Future single crystal neutron diffraction measurements in nonzero field will be needed to determine the potentially complicated magnetic phase diagram. Critical exponents computed from order parameter measurements obtained from neutron diffraction would also serve to tie the system to the XY model.

\begin{acknowledgments}

We acknowledge support from S. Calder during the neutron scattering experiment as well as Y. Hao, J.A.M. Paddison, and V.O. Garlea for support with magnetic structure analysis. This work was funded, in part, by the Gordon and Betty Moore Foundation EPiQS Initiative, Grant No. GBMF9070 to J.G.C (instrumentation development) and the Army Research Office, Grant No. W911NF-24-1-0234 (material characterization). A portion of this research used resources at the High Flux Isotope Reactor, a DOE Office of Science User Facility operated by the Oak Ridge National Laboratory. The beam time was allocated to HB-2A on proposal number 32050.1.

\end{acknowledgments}

\section*{DATA AVAILABILITY}

The data that support the findings of this article are openly available \cite{datastatement}.

\appendix

\section{Magnetization of the Kramers' Doublets}
\label{sec: Magnetization of the Kramers' Doublets}

The $D_{3h}$ local crystal field breaks up the originally sixfold degenerate manifold of $J=5/2$ states into three Kramers' doublets. The Hamiltonian describing the Zeeman splitting is given by:

\begin{equation}\label{Eq: Zeeman}
    \mathbf{H} = \mu_B \, \vec{H} \cdot \overset\leftrightarrow{g} \cdot \vec{\mathbf{J}}
\end{equation}

\noindent
where $\vec{\mathbf{J}}$ is the angular momentum operator (in units of $\hbar$), $\overset\leftrightarrow{g}$ is the g-tensor, and $\vec{H}$ is the magnetic field \cite{g-factor, g-factor_2}. In free-space, Eq. (\ref{Eq: Zeeman}) reduces to:

\begin{equation}\label{Eq: Zeeman52}
    \mathbf{H} = \mu_B g_J \left( H_x \mathbf{J}^x + H_y \mathbf{J}^y + H_z \mathbf{J}^z \right)
\end{equation}

\noindent
since the g-tensor is proportional to the identity operator. The Hamiltonian in Eq. (\ref{Eq: Zeeman52}) may be projected onto the $\ket{J=5/2, \pm m_J}$ Kramers' doublets of the $D_{3h}$ crystal field to yield:

\begin{gather}
    \mathbf{H} _{\pm 1/2} = \frac{3}{7} \begin{pmatrix}  H_z  & 3 \left( H_x - i H_y \right)\\ 3 \left( H_x + i H_y \right) & - H_z \end{pmatrix} \mu_B \\
    \mathbf{H} _{\pm 3/2} = \frac{9}{7} \begin{pmatrix}  H_z  & 0 \\ 0 & - H_z \end{pmatrix} \mu_B \\
    \mathbf{H} _{\pm 5/2} = \frac{15}{7} \begin{pmatrix}  H_z  & 0 \\ 0 & - H_z \end{pmatrix} \mu_B
\end{gather}

\noindent
where the Landé g-factor of $g_{J=5/2} = 6/7$ was used \cite{g-factor}. These projected hamiltonians my be rewritten in the form of Eq. (\ref{Eq: Zeeman}) as follows:

\begin{gather}
    \mathbf{H} _{\pm 1/2} = \mu_B \, \begin{pmatrix} H_x \\ H_y \\ H_z \end{pmatrix}^\intercal \begin{pmatrix}  \frac{18}{7} & 0 & 0 \\ 0 & \frac{18}{7} & 0 \\ 0 & 0 & \frac{6}{7} \end{pmatrix} \begin{pmatrix} \tilde{\mathbf{S}}^x \\ \tilde{\mathbf{S}}^y \\ \tilde{\mathbf{S}}^z \end{pmatrix} \\
    \mathbf{H} _{\pm 3/2} = \mu_B \, \begin{pmatrix} H_x \\ H_y \\ H_z \end{pmatrix}^\intercal \begin{pmatrix}  0 & 0 & 0 \\ 0 & 0 & 0 \\ 0 & 0 & \frac{18}{7} \end{pmatrix} \begin{pmatrix} \tilde{\mathbf{S}}^x \\ \tilde{\mathbf{S}}^y \\ \tilde{\mathbf{S}}^z \end{pmatrix} \\
    \mathbf{H} _{\pm 5/2} = \mu_B \, \begin{pmatrix} H_x \\ H_y \\ H_z \end{pmatrix}^\intercal \begin{pmatrix}  0 & 0 & 0 \\ 0 & 0 & 0 \\ 0 & 0 & \frac{30}{7} \end{pmatrix} \begin{pmatrix} \tilde{\mathbf{S}}^x \\ \tilde{\mathbf{S}}^y \\ \tilde{\mathbf{S}}^z \end{pmatrix}
\end{gather}

\noindent
where $\tilde{\mathbf{S}}^\alpha$ are the ``effective'' spin-$1/2$ angular momentum operators written in terms of the $2 \times 2$ Pauli operators as $\tilde{\mathbf{S}}^\alpha = \frac12 \boldsymbol{\sigma}^\alpha$ \cite{g-factor_2}. The components $g_\parallel$ (parallel to $\hat{z}$-axis) and $g_\perp$ (perpendicular to $\hat{z}$-axis) may then be read off of the g-tensor in these hamiltonians for each Kramers' doublet.

The Zeeman hamiltonian has been projected onto these Kramers' doublets such that each doublet effectively functions as a spin-$1/2$ system with eigenenergies of $E_\pm = \pm \frac12 g_{\parallel, \perp} \mu_B \, H$ depending on the direction of the applied magnetic field to the $\hat{z}$-axis set by the principle rotation axis of the crystal field. The expected magnetization of each Kramers' doublet then follows the standard spin-$1/2$ form with $\mu \equiv \frac12 g_{\parallel, \perp} \mu_B$ shown in Eq. (\ref{Eq: M12}).

\begin{equation}\label{Eq: M12}
    M = \frac12 g_{\parallel, \perp} \mu_B \, \tanh{\left(  \frac{g_{\parallel, \perp} \mu_B \, H}{2 k_B T}  \right)}
\end{equation}

The low temperature magnetization may be fit by Eq. (\ref{Eq: M12}) both as a function of temperature, and as a function of field to extract $g_{\parallel, \perp}$ for the ground state Kramers' doublet under the assumption that this is the only populated energy level at such low temperatures.

\begin{figure}[H]
	\centering 
	\includegraphics[width=1\linewidth]{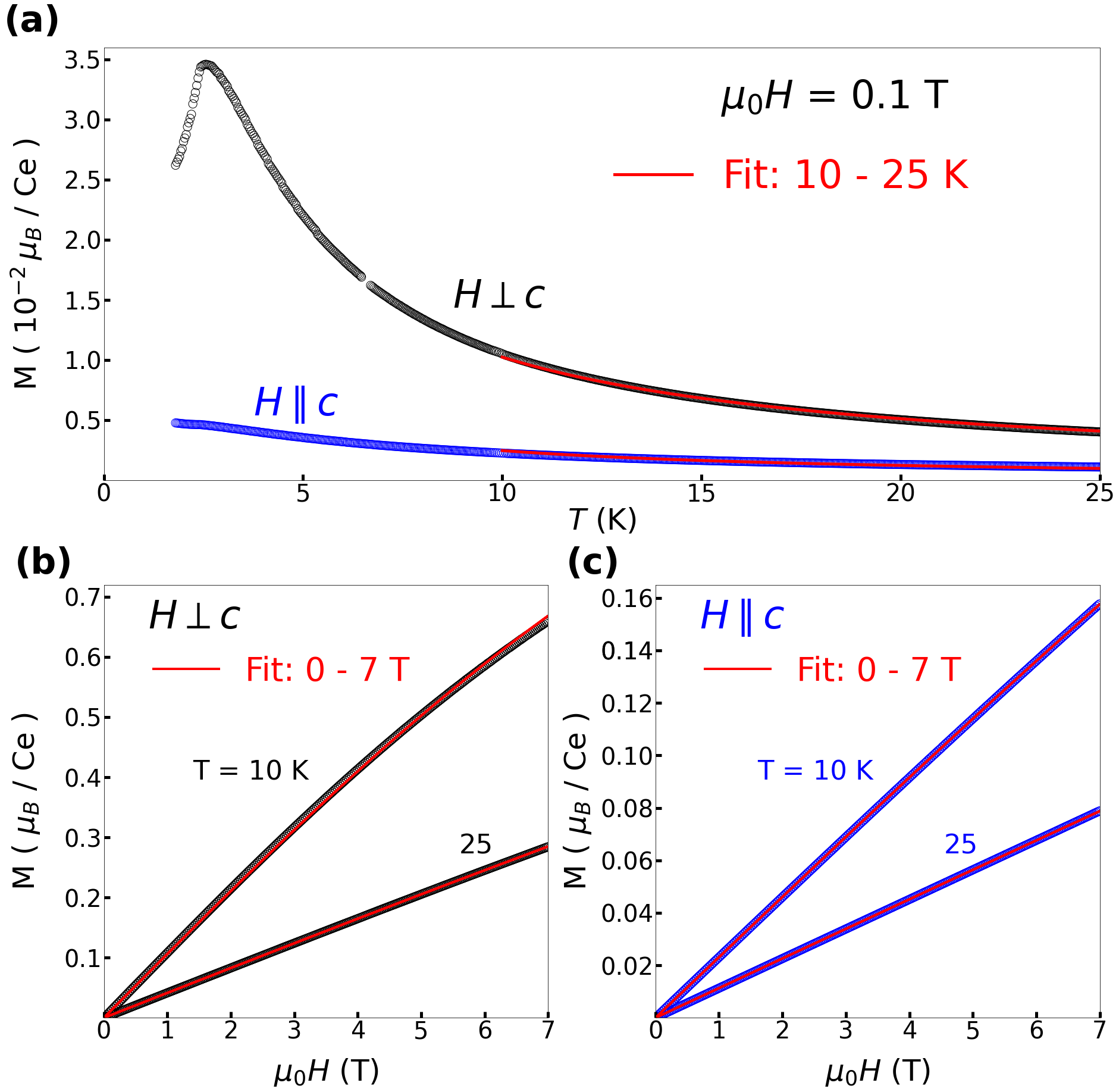}
	\caption{Magnetization of crystal \textit{A} measured as a function of \textbf{(a)} temperature at fixed field ($0.1 \, \text{T}$), and as a function of \textbf{(b, c)} field at fixed temperature ($10$ and $25$ K). Measurements were taken with field aligned perpendicular to the \textit{c}-axis (black) and parallel to the \textit{c}-axis (blue). The dashed red line shows the fit of the magnetization using Eq. (\ref{Eq: M12}).}
	\label{fig:gFit}
\end{figure}

Figure \ref{fig:gFit} shows this fit for both orientations of the crystal relative to the magnetic field. Note that due to the orientation of the CEF potential in Eq. (\ref{Eq: TTP}) relative to the crystal structure, $g_\parallel$ is extracted from the measurement when field is parallel to the $c$-axis, and $g_\perp$ is extracted from the measurement when field is perpendicular to the $c$-axis. The g-factors for the ground state extracted by this fit are presented in Table \ref{T: g} along with the values predicted for the ground state spanned by $\ket{\psi_{\pm1/2}}$.

\begin{table}[H]
\caption{\label{T: g}Ground state g-factors extracted for Ce\textsubscript{2}SnS\textsubscript{5} crystal $A$ by fitting magnetization vs. temperature at $0.1 \, \text{T}$ from $5-10\, \text{K}$ and magnetization vs. field at $5$ and $10$ K from $0-7 \, \text{T}$. The uncertainties extracted from the fits are all between three to five orders of magnitude smaller than the g-factors themselves.}
\begin{ruledtabular}
\begin{tabular}{p{3.1cm} | P{1.8cm} P{1.8cm} P{1.8cm}}
   &  $g_\perp$ &  $g_\parallel$ & $g_\perp/g_\parallel$   \\ [0.6ex] 
  \hline
  Predicted for $\ket{\psi_{\pm1/2}}$  & 18/7 & 6/7 & 3 \\ [0.6ex] 
  \hline
  M vs. H at $10$ K & $2.52$ & $1.17$ & $2.147$ \\ [1ex] 
  M vs. H at $25$ K & $2.48$ & $1.30$ & $1.915$ \\ [1ex] 
  M vs. T at $0.1$ T & $2.47$ & $1.21$ & $2.035$ \\ [0.6ex] 
\end{tabular}
\end{ruledtabular}
\end{table}

\section{Fitting the Heat Capacity}
\label{sec: Fitting the Heat Capacity}

The heat capacity of Ce\textsubscript{2}SnS\textsubscript{5} is modeled extremely well by phonon mode contributions in the region well above the magnetic transition. This phonon contribution serves as a background at low temperatures such that the magnetic contributions to heat capacity may be isolated and the magnetic heat capacity computed. The phonon heat capacity of the system may be fit via a combined anharmonic Einstein and Debye model as given by \cite{Martin_1991}:

\begin{multline}\label{eq:HC}
    \frac{C}{Nk_B} = \sum_{j \in optic} \frac{1}{1 - C_j T} \frac{\epsilon_j^2 \, e^{\epsilon_j/T}}{\left( e^{\epsilon_j/T} - 1\right)^2}\\
    + \frac{9}{1 - C_D T} \left( \frac{T}{\theta_D} \right)^3 \int_0^{\theta_D/T} \frac{x^4 e^x \, dx}{\left( e^x - 1 \right)^2}
\end{multline}

\noindent
where $\theta_D$ is the Debye temperature. The summation in the Einstein term is over all optic modes and $\epsilon_j = \hbar \omega_j/k_B$ is the energy (in units of temperature) of the optic mode with frequency $\omega_j$. The anharmonicities of the optic modes and the Debye term are given by $C_j$ and $C_D$ respectively; in this model, the oscillator frequency $\omega_j$ is given by \cite{Martin_1991}:

\begin{equation}
    \omega_j = \omega_0 \left( 1 - C_j T \right)
\end{equation}

\noindent
where $\omega_0$ is the harmonic oscillator frequency and $C_j$ is the anharmonicity. Based on Eq. (\ref{eq:HC}), the phonon term of the heat capacity may be fit using the following model:

\begin{multline}\label{eq:model}
    C_{ph} \sim 3 \alpha (1 + C_E T) \frac{\epsilon^2 \, e^{\epsilon/T}}{\left( e^{\epsilon/T} - 1\right)^2}\\
    + 9 (1- \alpha) (1 + C_D T) \left( \frac{T}{\theta_D} \right)^3 \int_0^{\theta_D/T} \frac{x^4 e^x \, dx}{\left( e^x - 1 \right)^2}
\end{multline}

\noindent
where $0 \leq \alpha \leq 1$ is an interpolation parameter between the Einstein and Debye terms. In Eq. (\ref{eq:model}), it is assumed that there is only one branch of isotropic optic modes in three dimensions, and that the anharmonicities are small such that $\left( 1 - C_j T \right)^{-1} \approx \left( 1 + C_j T \right)$. The inclusion of the anharmonicities allows the model to remain valid up to higher temperatures whereat the anharmonic effects become evident.

\begin{figure}[H]
	\centering 
	\includegraphics[width=1\linewidth]{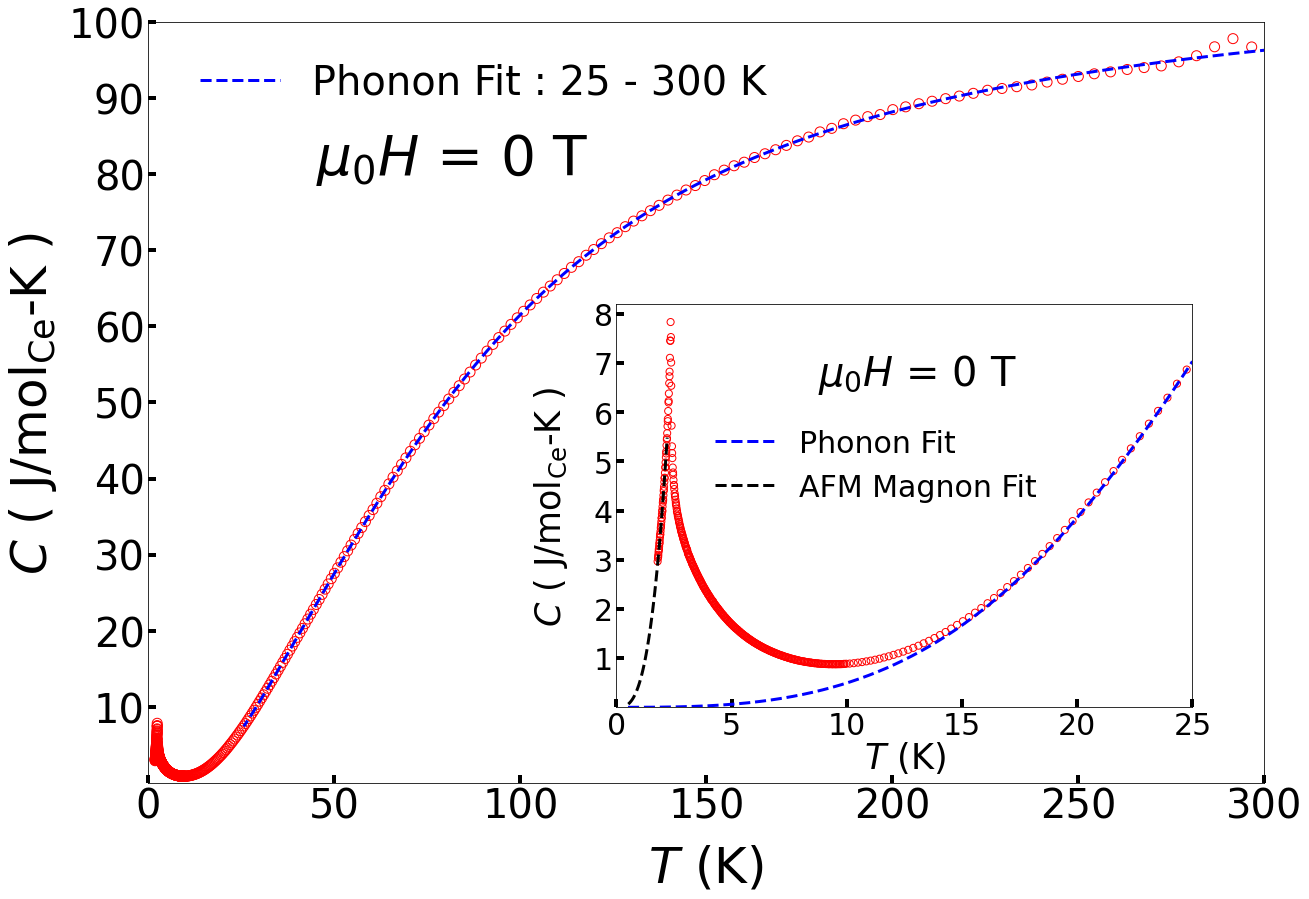}
	\caption{Zero-field heat capacity of Ce\textsubscript{2}SnS\textsubscript{5} with the phonon fit (between $25 \, \text{K}$ and $300 \, \text{K}$) drawn in blue. The inset shows the phonon fit extrapolated below $25 \, \text{K}$ (dashed blue line) as well as the AFM magnon fit (dashed black line) extrapolated below $1.8 \, \text{K}$.}
	\label{fig:ZeroFieldFit}
\end{figure}

The heat capacity data was fit between $25 \, \text{K}$ and $300 \, \text{K}$ (well above $T_N = 2.4 \, \text{K}$) to yield a Debye temperature of $\theta_D = 198 \pm 7 \, \text{K}$, and an optic mode energy of $\epsilon = 348 \pm 21 \, \text{K}$ with an interpolation parameter of $\alpha = 0.51 \pm 0.05$. 
This fit along with the heat capacity data is shown in Fig. \ref{fig:ZeroFieldFit}. Note that the apparent feature in the data near $300 \, \text{K}$ is a measurement anomaly likely due to freezing of grease in the cryostat upon cooling. 

The model in Eq. (\ref{eq:model}) is applied in the high-temperature, phonon-dominated region of the measurement such that it may provide an appropriate extrapolated background for analysis of the low temperature regime where magnetic interactions become relevant. The inset of Fig. \ref{fig:ZeroFieldFit} shows the extrapolated high temperature phonon fit in blue; this fit remains smooth and does not cross the data such that it provides a high quality background.

The dashed black line in the inset of Fig. \ref{fig:ZeroFieldFit} is the antiferromagnetic (AFM) magnon fit below the Neel temperature. This was fit via a $C_M \sim T^\alpha$ power law between $1.8 \, \text{K}$ and $2.2 \, \text{K}$ to yield an exponent of $\alpha = 3.03 \pm 0.02$ which agrees well with the prediction for three-dimensional AF magnons. In three dimensions, magnons are expected to exhibit a $C \sim T^{3/2}$ heat capacity in a ferromagnetically ordered system. Since an antiferromagnetic lattice can be broken up into a bipartite sum of ferromagnetic sublattices, an AF lattice with two such sublattices exhibit a $C \sim T^{3}$ heat capacity. This simple functional form was used to extrapolate the heat capacity from $1.8 \, \text{K}$ to zero to allow for the magnetic entropy to be computed.

\section{The Schottky Anomaly}
\label{sec: The Schottky Anomaly}

\begin{figure}[H]
	\centering 
	\includegraphics[width=1\linewidth]{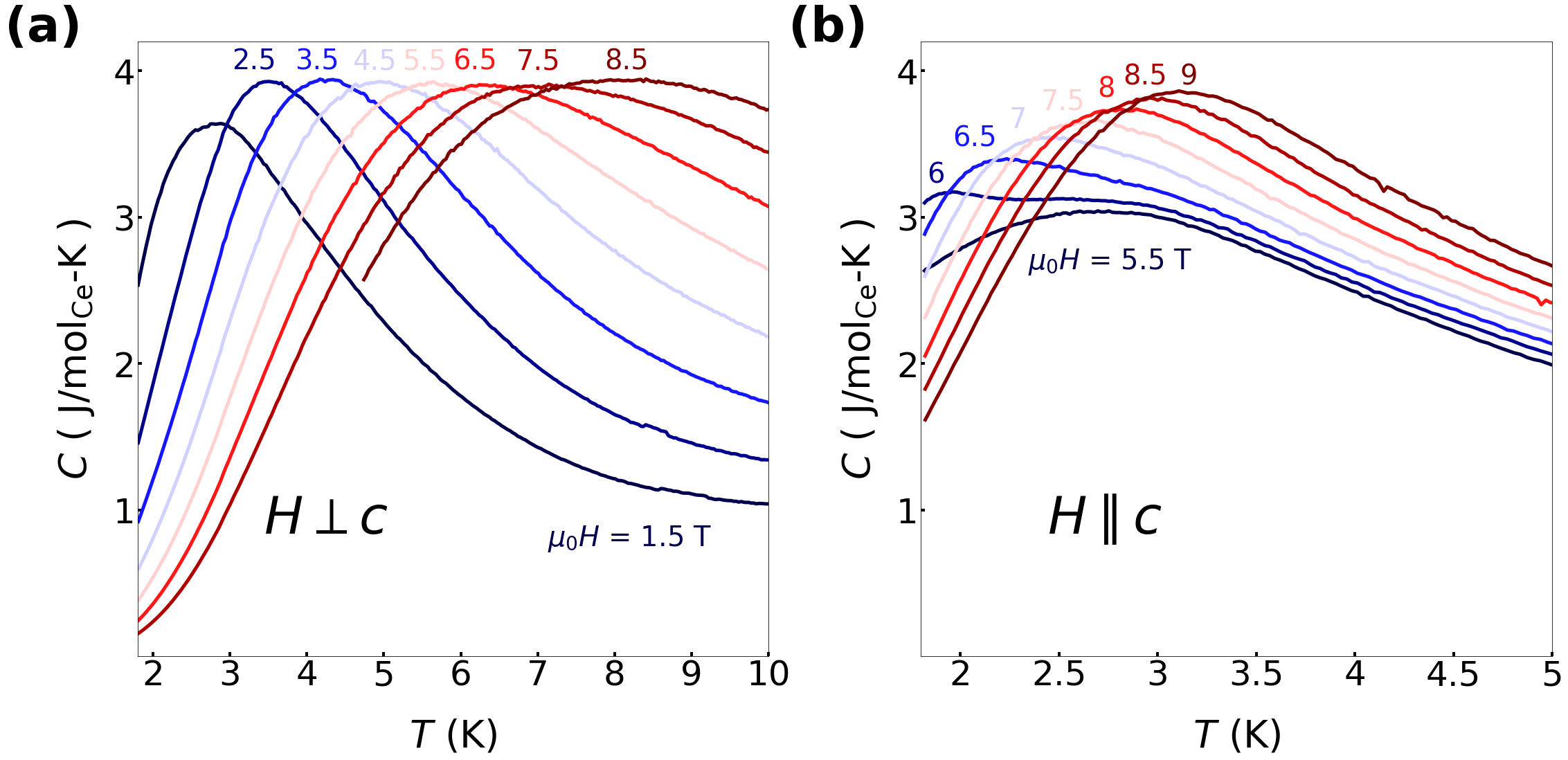}
	\caption{Evolution of the Schottky anomaly in various applied magnetic fields. \textbf{(a)} Heat capacity of crystal \textit{B} with field aligned perpendicular to the \textit{c}-axis. \textbf{(b)} Heat capacity of crystal \textit{C} with field aligned parallel to the \textit{c}-axis.}
	\label{fig:Schottky}
\end{figure}

The Ce\textsubscript{2}SnS\textsubscript{5} system exhibits a pronounced Schottky anomaly as increasing magnetic field strength is applied to the sample. For fields perpendicular to the $c$-axis, this occurs roughly above $\mu_0 H = 1 \, \text{T}$. For fields parallel to the $c$-axis, this occurs roughly above $\mu_0 H = 5.5 \, \text{T}$. Fig. \ref{fig:Schottky} shows the evolution of the Schottky anomaly in field for both crystal orientations.

The Schottky anomaly was unable to be fit at such high fields due to the number of relevant free parameters. Due to the crystal field distortion, the relevant g-factors differ from those computed in Eqs. (\ref{Eq: g1/2} - \ref{Eq: g5/2}) and the exact values become important at high fields which necessitates their being fit as free parameters. Due to the lack of a zero-field Schottky anomaly, the two zero-field energy level splittings also need to be fit at high field. The resulting five-parameter did not converge to consistent values for measurements at different fields for the same crystal orientation.

\section{Fitting the Magnetic Reflection Peaks}
\label{sec: Fitting the Magnetic Reflection Peaks}

Powder neutron diffraction spectra of Ce\textsubscript{2}SnS\textsubscript{5} reveal magnetic reflection peaks under $T_N = 2.4 \, \text{K}$. The following outlines our procedure for fitting selected individual magnetic reflection peaks and refining the magnetic structure which best fits the low-temperature spectrum.

\subsection{Fitting Selected Reflection Peaks}
\label{sec: Fitting Selected Reflection Peaks}

The powder neutron diffraction spectra taken under $T_N = 2.4 \, \text{K}$ show several magnetic reflections which appear at $q$-vectors distinct from the nuclear reflections. To fit these selected magnetic reflection peaks, a $4 \, \text{K}$ background spectrum was subtracted from the spectra taken under $T_N$, and each peak was fit with a Lorentzian lineshape:

\begin{equation}\label{Eq: L}
    f(q) = A \, \frac{\gamma^2}{\gamma^2 + (q - q_0)^2} + f_0
\end{equation}

\begin{figure}[H]
	\centering 
	\includegraphics[width=1\linewidth]{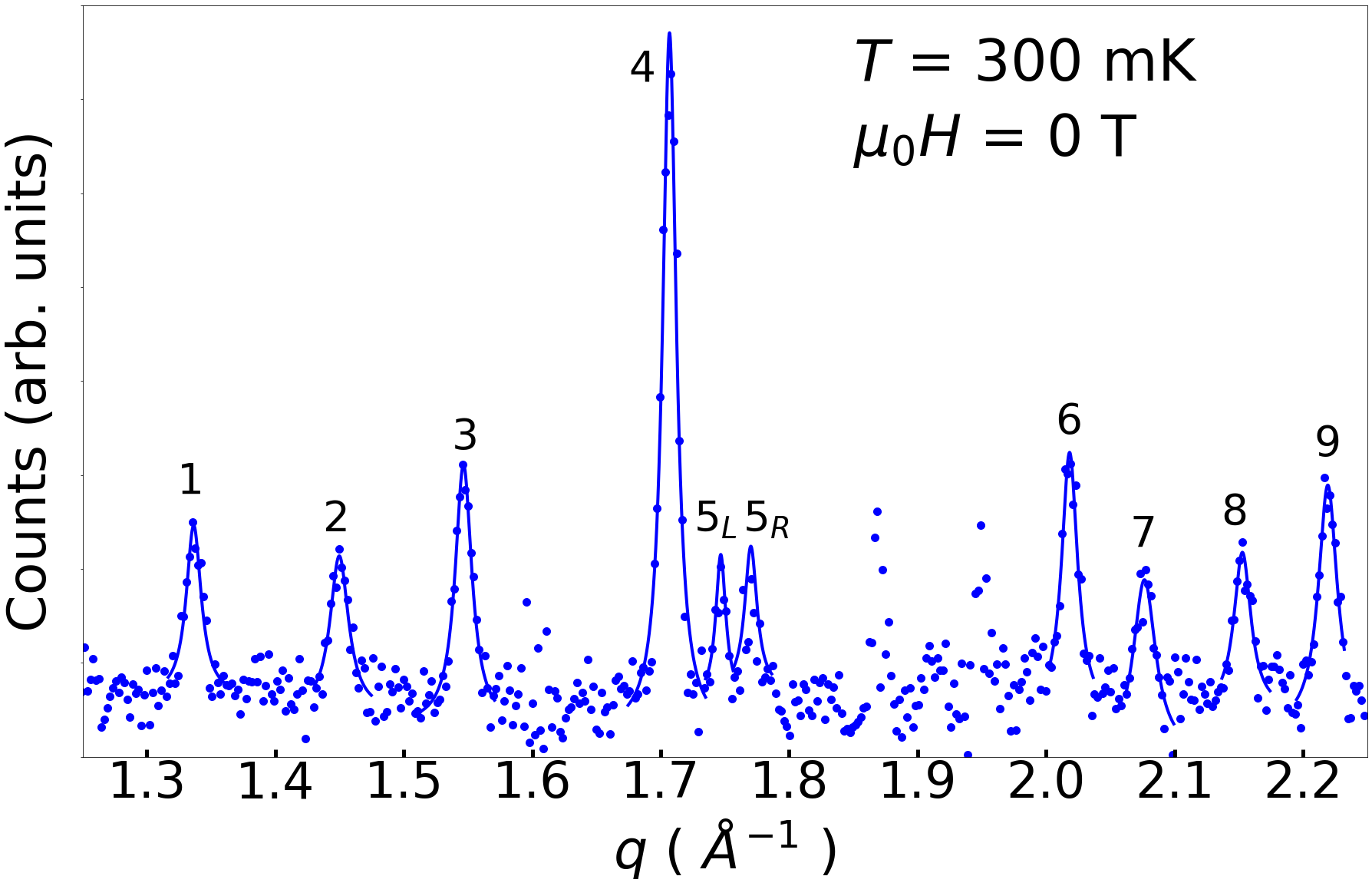}
	\caption{\textbf{(a)} Zero-field, background-subtracted powder neutron diffraction spectrum ($\lambda = 2.41 \, \text{\AA}$) of Ce\textsubscript{2}SnS\textsubscript{5} at $300 \, \text{mK}$. The Lorentzian fits of the labeled selected magnetic reflection peak are shown as solid blue lines}
	\label{fig:13peaks}
\end{figure}

\begin{figure}[H]
	\centering 
	\includegraphics[width=1\linewidth]{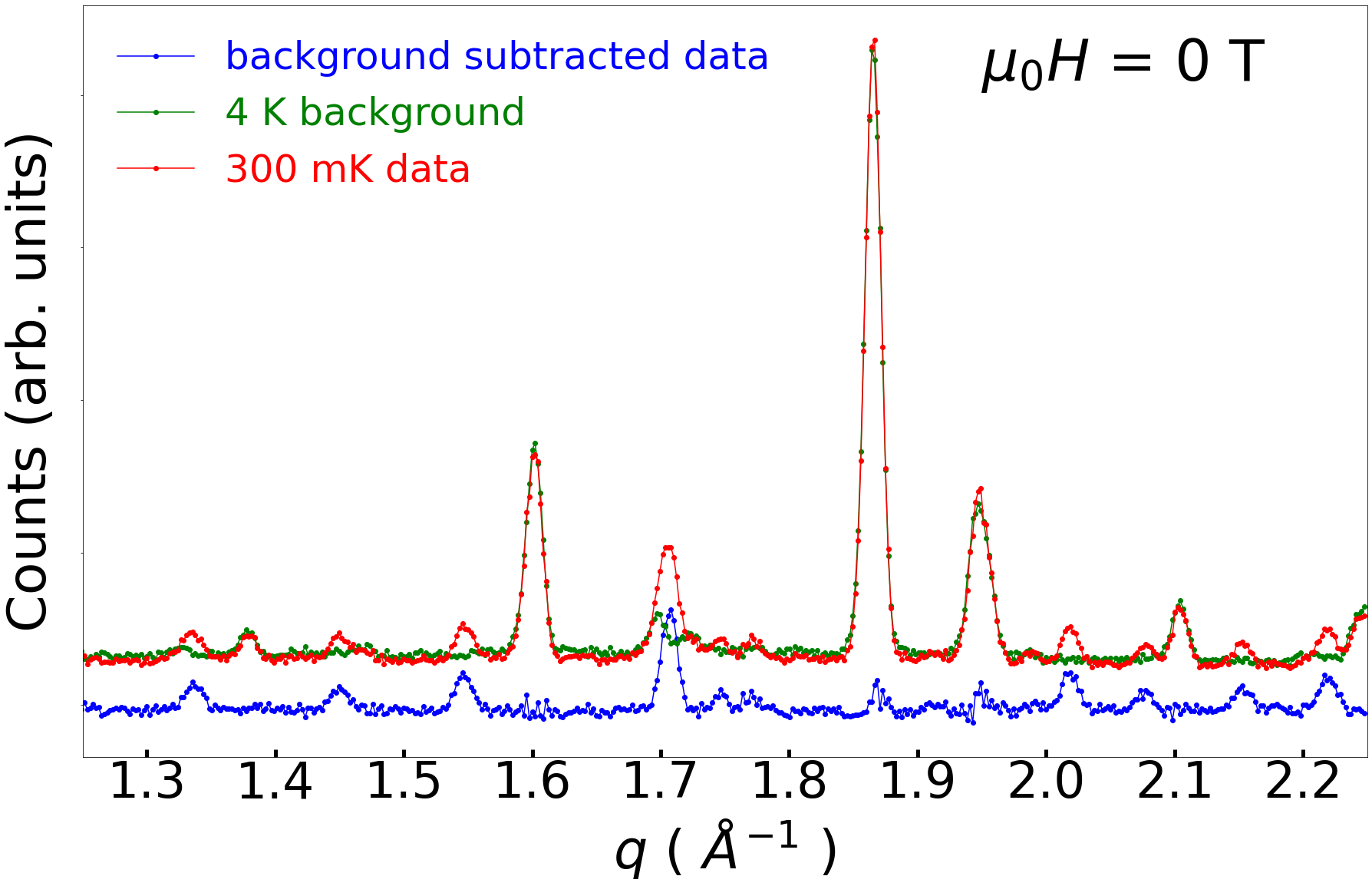}
	\caption{Zero-field powder neutron diffraction spectra ($\lambda = 2.41 \, \text{\AA}$) of Ce\textsubscript{2}SnS\textsubscript{5} taken at $4 \, \text{K}$ (green) and $300 \, \text{mK}$ (red). The background subtracted $300 \, \text{mK}$ spectrum is shown in blue.}
	\label{fig:403}
\end{figure}

\bigskip
\noindent
where $A$ is the peak amplitude, $\gamma$ is the half width at half maximum (HWHM), $q_0$ is the peak location, and $f_0$ is a constant offset. Using the k-search program in the FullProf software suite \cite{FullProf}, the values of $q_0$ for ten magnetic reflections distinct from the nuclear reflections were used to determine that the best fitting commensurate propagation vector associated with the magnetic unit cell was $\vec{q}=(1/3,0,0)$. Figure \ref{fig:13peaks} shows the background-subtracted diffraction spectrum taken at $300 \, \text{mK}$ with the Lorentzian fits of selected peaks plotted over the data, and Fig. \ref{fig:403} shows the $4 \, \text{K}$ background and $300 \, \text{mK}$ raw data spectrum plotted together. The peaks located near $q=1.87 \, \text{\AA}$ and $q=1.95 \, \text{\AA}$ in Fig. \ref{fig:13peaks} may be attributed to satellite peaks from the $\vec{q}=(1/3,0,0)$ propagation vector which lie near the nuclear peak positions, magnetic reflections from the $\vec{q}=(0,0,0)$ propagation vector, and lattice contraction shifting the locations of the nuclear peaks relative to the background spectrum.

Each data point in the raw diffraction spectra was assumed to follow a Poisson distribution such that the standard deviation of a data point with $C$ counts is $\sigma_C = \sqrt{C}$. Let the $i$\textsuperscript{th} data point of the full diffraction spectrum taken under $T_N$ be denoted by $D_i$, and let the corresponding data point of the background spectrum taken at $T=4 \, \text{K}$ be denoted by $B_i$ such that:

\begin{gather}
    D_i \sim Pois\big( D_i \big)\\
    B_i \sim Pois\big( B_i \big)
\end{gather}

Let the $i$\textsuperscript{th} data point of the piece of the diffraction spectrum under $T_N$ corresponding to the magnetic reflections be denoted by $M_i$ such that:

\begin{equation}
    M_i = D_i - B_i
\end{equation}

Clearly, $D_i$ and $B_i$ are dependent random variables, but they share a common, additive random contribution (since $D_i = M_i + B_i$). This yields the result that the random variable $M_i$ follows a Skellam distribution with parameters $\mu_1 = D_i$ and $\mu_2 = B_i$, \textit{i.e.}:

\begin{gather}
    M_i \sim Skellam\big( D_i,  B_i \big)
\end{gather}

As the variance of $Skellam(\mu_1, \mu_2)$ is given by $\mu_1 + \mu_2$, the standard deviation of $M_i$ is:

\begin{equation}
    \sigma_{M_i} = \sqrt{T_i + B_i}
\end{equation}

A least-squares fit of the Lorentzian lineshape described in Eq. (\ref{Eq: L}) was then performed on the background subtracted dataset around each magnetic reflection peak to extract its position $q_0$. The standard deviation of the $q_0$ extracted from the fit was obtained by taking the square root of the corresponding diagonal element of the covariance matrix of the fit. Tables \ref{T: N} and \ref{T: N5} shows the $q_0$ values extracted by this fit as well as the corresponding standard deviation $\sigma_{q_0}$ for each of these magnetic reflection peaks.

The temperature dependent peak shift $\Delta q_0$ is defined as:

\begin{equation}
    \Delta q_0 = q_0 - q_0^*
\end{equation}

\noindent
where $q_0^*$ is the value of the peak position $q_0$ obtained from the $T = 2 \, \text{K}$ spectrum. The variance of $\Delta q_0$ is then given by:

\begin{equation}
    \text{Var} \big( \Delta q_0 \big) = \text{Var} \big( q_0 \big) + \text{Var} \big( q_0^* \big) - 2 \, \text{Cov} \big( q_0, \, q_0^* \big)
\end{equation}

\noindent
where it is assumed that $\text{Cov}( q_0, \, q_0^* ) = 0$ except for when $q_0 = q_0^*$ whereat $\text{Cov}( q_0, \, q_0 ) = Var(q_0)$. The widths of the error bars in Fig. \ref{fig:MagRefEv}(a) are given by $2 \sqrt{\text{Var} ( \Delta q_0 )}$ for each value of $\Delta q_0$.

\begin{table*}
\caption{\label{T: N}Peak locations ($q_0$) and standard deviations ($\sigma_{q_0}$) extracted from Lorentzian fits of the magnetic reflection peaks of Ce\textsubscript{2}SnS\textsubscript{5} obtained from zero-field powder neutron diffraction at several temperatures under $T_N = 2.4 \, \text{K}$. The peak numbers correspond to those in Fig. \ref{fig:13peaks}.}
\begin{ruledtabular}
\begin{tabular}{cccc|cccc|cccc}
  Peak & $T$ (K) &  $q_0$ ($\text{\AA}^{-1}$) &  $\sigma_{q_0}$ ($10^{-3} \text{\AA}^{-1}$)$\quad$ & Peak & $T$ (K) &  $q_0$ ($\text{\AA}^{-1}$) &  $\sigma_{q_0}$ ($10^{-3} \text{\AA}^{-1}$)$\quad$ & Peak & $T$ (K) &  $q_0$ ($\text{\AA}^{-1}$) &  $\sigma_{q_0}$ ($10^{-3} \text{\AA}^{-1}$) \\ [0.8ex] 
  \hline
  \textbf{1} & 2.0 & 1.3525 & 1.1835 & \textbf{4} & 2.0 & 1.7046 & 0.4102 & \textbf{8} & 2.0 & 2.1619 & 1.2449\\ [0.6ex] 
    & 1.8 & 1.3494 & 0.8948 &   & 1.8 & 1.7050 & 0.4814 &   & 1.8 & 2.1618 & 1.4510 \\ [0.6ex] 
    & 1.5 & 1.3453 & 0.7693 &   & 1.5 & 1.7054 & 0.3473 &   & 1.5 & 2.1591 & 0.9034 \\ [0.6ex] 
    & 1.2 & 1.3366 & 1.0482 &   & 1.2 & 1.7071 & 0.3306 &   & 1.2 & 2.1539 & 1.0455 \\ [0.6ex] 
    & 0.8 & 1.3351 & 1.0573 &   & 0.8 & 1.7068 & 0.3038 &   & 0.8 & 2.1539 & 1.2701 \\ [0.6ex] 
    & 0.5 & 1.3363 & 0.7847 &   & 0.5 & 1.7070 & 0.2899 &   & 0.5 & 2.1545 & 0.9207 \\ [0.6ex] 
    & 0.3 & 1.3365 & 0.5752 &   & 0.3 & 1.7069 & 0.1773 &   & 0.3 & 2.1529 & 0.6095 \\ [0.6ex] 
  \hline
  \textbf{2} & 2.0 & 1.4628 & 1.2474 & \textbf{6} & 2.0 & 2.0226 & 0.7943 & \textbf{9} & 2.0 & 2.2142 & 1.0023\\ [0.6ex] 
    & 1.8 & 1.4622 & 1.4791 &   & 1.8 & 2.0220 & 0.6803 &   & 1.8 & 2.2148 & 0.7947 \\ [0.6ex] 
    & 1.5 & 1.4581 & 1.4895 &   & 1.5 & 2.0203 & 0.6518 &   & 1.5 & 2.2149 & 0.7677 \\ [0.6ex] 
    & 1.2 & 1.4519 & 0.6417 &   & 1.2 & 2.0178 & 0.5832 &   & 1.2 & 2.2196 & 0.7110 \\ [0.6ex] 
    & 0.8 & 1.4503 & 0.8340 &   & 0.8 & 2.0183 & 0.4941 &   & 0.8 & 2.2178 & 0.7361 \\ [0.6ex] 
    & 0.5 & 1.4518 & 0.7701 &   & 0.5 & 2.0181 & 0.6343 &   & 0.5 & 2.2197 & 0.6646 \\ [0.6ex] 
    & 0.3 & 1.4498 & 0.6504 &   & 0.3 & 2.0184 & 0.3656 &   & 0.3 & 2.2194 & 0.4627 \\ [0.6ex] 
  \hline
  \textbf{3} & 2.0 & 1.5361 & 0.7864 & \textbf{7} & 2.0 & 2.0885 & 1.2205 & \textbf{10} & 2.0 & 3.5273 & 1.6729\\ [0.6ex] 
    & 1.8 & 1.5376 & 0.8419 &   & 1.8 & 2.0871 & 2.0735 &   & 1.8 & 3.5295 & 1.7698 \\ [0.6ex] 
    & 1.5 & 1.5397 & 0.5757 &   & 1.5 & 2.0824 & 0.8781 &   & 1.5 & 3.5291 & 0.7773 \\ [0.6ex] 
    & 1.2 & 1.5467 & 0.5495 &   & 1.2 & 2.0765 & 1.0520 &   & 1.2 & 3.5300 & 0.8236 \\ [0.6ex] 
    & 0.8 & 1.5461 & 0.7012 &   & 0.8 & 2.0769 & 1.0830 &   & 0.8 & 3.5302 & 0.7955 \\ [0.6ex] 
    & 0.5 & 1.5475 & 0.6549 &   & 0.5 & 2.0767 & 1.1134 &   & 0.5 & 3.5295 & 1.2196 \\ [0.6ex] 
    & 0.3 & 1.5465 & 0.3876 &   & 0.3 & 2.0766 & 0.6701 &   & 0.3 & 3.5295 & 0.5814 \\ [0.6ex] 
\end{tabular}
\end{ruledtabular}
\end{table*}

\begin{table*}
\caption{\label{T: N5}Peak locations ($q_0$) and standard deviations ($\sigma_{q_0}$) extracted from Lorentzian fits of the split peak $\mathbf{5_L}$/$\mathbf{5_R}$ in Fig. \ref{fig:13peaks}. Note that this magnetic reflection peak bifurcates under $1.5 \, \text{K}$.}
\begin{ruledtabular}
\begin{tabular}{ccc}
  $T$ (K) &  $q_0$ ($\text{\AA}^{-1}$) &  $\sigma_{q_0}$ ($10^{-3} \text{\AA}^{-1}$)$\quad$ \\ [0.8ex] 
  \hline
    2.0 & $1.7576$ & $1.6542$ \\ [0.6ex] 
    1.8 & $1.7652$ & $1.4265$ \\ [0.6ex] 
    1.5 & $1.7662$ & $2.9768$ \\ [0.6ex] 
    1.2 & $1.7473 \qquad \qquad 1.7687$ & $1.1387 \qquad \qquad 1.0384$\\ [0.6ex] 
    0.8 & $1.7455 \qquad \qquad 1.7673$ & $1.1534 \qquad \qquad 0.9714$\\ [0.6ex] 
    0.5 & $1.7481 \qquad \qquad 1.7709$ & $1.3322 \qquad \qquad 1.3299$\\ [0.6ex] 
    0.3 & $1.7468 \qquad \qquad 1.7702$ & $0.6484 \qquad \qquad 1.0341$\\ [0.6ex] 

\end{tabular}
\end{ruledtabular}
\end{table*}

\subsection{Fitting The Magnetic Structure}
\label{sec: Fitting The Magnetic Structure}

The low-temperature, commensurate magnetic structure was fitted using a Rietveld refinement implemented using the FullProf software suite \cite{FullProf}. The low-temperature crystal structure was fit using the $4 \, \text{K}$ data, and lattice parameters, profile parameters, and zero offset were fixed to these best fit values for the magnetic structure refinement. A list of eight maximal likelihood Shubnikov groups compatible with a $\vec{q}=(1/3,0,0)$ propagation vector was prepared using MAXMAGN in the Bilbao crystallographic server \cite{PerezMato2015}. For all of these Shubnikov groups, the cerium position splits into three symmetry independent sites, and the magnetic moments lie either within the \textit{ab}-plane ($M_z = 0$) or along the \textit{c}-axis ($M_x, M_y = 0$). During refinement, the magnitude of the cerium moment was additionally constrained to be equal on each site. 

Subject to these constraints, the background-subtracted $300 \, \text{mK}$ spectrum was fit to each of these maximal likelihood Shubnikov groups. From these refinements, the best fitting magnetic structure has moments constrained to lie within the \textit{ab}-plane and belongs to the Shubnikov group \textit{Pb’a’m’} (MSG 55.359) with a goodness of fit of $\chi^2 = 1.22$. Note that a refinement of MSG 55.358 yields a relatively similar goodness of fit of $\chi^2 = 1.30$, but the fit is noticeably worse. MSG 55.359 is additionally compatible with an additional $\vec{q} = (0,0,0)$ irreducible representation which suggests a 2-\textit{q} magnetic structure. This conclusion is supported by the inability to fit the neutron diffraction data with irreducible representations from $\vec{q}=(1/3,0,0)$ alone \cite{Wills2000SARAh}. Magnetic structure parameters for the low-temperature commensurate phase are presented in Table \ref{T:Ce2SnS5_magnetic}.


\begin{table*}
\caption{Magnetic structure parameters of Ce$_2$SnS$_5$ for the low-temperature commensurate phase.}
\label{T:Ce2SnS5_magnetic}
\begingroup
\renewcommand{\arraystretch}{1.40}
\begin{tabular*}{\textwidth}{@{\extracolsep{\fill}} ll}
\hline\hline
Parent space group & \textit{Pbam} (No.~55) \\[8pt]\hline
Propagation vectors & $(\tfrac13,0,0)$ and $(0,0,0)$ \\[8pt]\hline
Transformation (parent $\to$ magnetic basis) & $(3a,b,c;0,0,0)$ \\[8pt]\hline
Shubnikov group & \textit{Pb'a'm'} (MSG 55.359) \\[8pt]\hline
Transformation (magnetic basis $\to$ standard setting) & $(a,b,c;\tfrac16,0,0)$ \\[8pt]\hline
Magnetic unit cell parameters (\AA) &
$a=23.5625$, $b=11.2177$, $c=3.95501$, $\alpha=\beta=\gamma=90^{\circ}$ \\[8pt]\hline
Shubnikov group symmetry operations &
\begin{tabular}[t]{@{}l@{}}
$(x,y,z; +1)$\\
$(x,y,-z; -1)$\\
$(-x+\tfrac13,-y,-z; -1)$\\
$(-x+\tfrac13,-y,z; +1)$\\
$(x+\tfrac12,-y+\tfrac12,-z; +1)$\\
$(x+\tfrac12,-y+\tfrac12,z; -1)$\\
$(-x+\tfrac56,y+\tfrac12,-z; +1)$\\
$(x+\tfrac56,y+\tfrac12,z; -1)$
\end{tabular} \\[8pt]\hline
Positions of magnetic atoms &
\begin{tabular}[t]{@{}l@{}}
Ce1\_1: $(0.02324,\,0.33084,\,0.50000)$\\
Ce1\_2: $(0.97676,\,0.66916,\,0.50000)$\\
Ce1\_3: $(0.14342,\,0.83084,\,0.50000)$
\end{tabular} \\[8pt]\hline
Positions of non-magnetic atoms &
\begin{tabular}[t]{@{}l@{}}
Sn1\_1: $(0,0,0)$\\
Sn1\_2: $(\tfrac16,\,\tfrac12,\,0)$\\
S1\_1: $(0.06217,\,0.07225,\,0.50000)$\\
S1\_2: $(0.93783,\,0.92775,\,0.50000)$\\
S1\_3: $(0.10450,\,0.57225,\,0.50000)$\\
S2\_1: $(0.11768,\,0.29875,\,0.00000)$\\
S2\_2: $(0.88232,\,0.70125,\,0.00000)$\\
S2\_3: $(0.04898,\,0.79875,\,0.00000)$\\
S3\_1: $(0,\,0.50000,\,0.00000)$\\
S3\_2: $(\tfrac16,\,0,\,0)$
\end{tabular} \\[8pt]\hline
Magnetic moments ($\mu_{\mathrm B}$) &
\begin{tabular}[t]{@{}l@{}}
Ce1\_1: $(0.20(3),\,1.250(10),\,0)$; $|\mathbf m|=1.265$\\
Ce1\_2: $(0.84(2),\,-0.946(19),\,0)$; $|\mathbf m|=1.265$\\
Ce1\_3: $(-0.35(3),\,-1.215(11),\,0)$; $|\mathbf m|=1.265$
\end{tabular} \\
\hline\hline
\end{tabular*}
\endgroup
\end{table*}


\clearpage
\bibliography{apssamp}

\end{document}